\input{psfig}

\documentstyle[12pt,aaspp4]{article}
\def\arcsec {$^{\prime \prime}$}
\def\arcmin {$^{\prime}$}

\def\etal   {{\it et~al.\/}}

\def\hbeta  {H$\beta$}
\def\halpha  {H$\alpha$}
\def\HII    {H~{\sc {II}}}
\def\kms    {~km~s$^{-1}$}

\begin{document}

\title{On Measuring Nebular Chemical Abundances in Distant
Galaxies Using Global Emission Line Spectra}
\author{Henry A. Kobulnicky\footnote{Hubble Fellow}\footnote{Visiting Astronomer, Kitt Peak National Observatory, National Optical Astronomy
     Observatories, which is operated by the Association of Universities for Research in
     Astronomy, Inc. (AURA) under cooperative agreement with the National Science
     Foundation. }\footnote{Visiting Astronomer, Cerro Tololo Observatory, National Optical Astronomy
     Observatories, which is operated by the Association of Universities for Research in
     Astronomy, Inc. (AURA) under cooperative agreement with the National Science
     Foundation. }\footnote{Visiting Astronomer, German-Spanish Astronomical Center, Calar Alto, operated 
	jointly by the Max-PlankInstitut f\"ur Astronomie and the Spanish National
	Commission for Astronomy}}
\affil{University of California, Santa Cruz \\ Lick Observatory/Board
of Studies in Astronomy \\ 
Santa Cruz, CA, 95064  \\ 
Electronic Mail: chip@ucolick.org}
\authoremail{chip@ucolick.org}
\author{Robert C. Kennicutt, Jr. and James L. Pizagno}
\affil{University of Arizona \\ Steward Observatory \\ 
Tucson, AZ 85721  \\ 
Electronic Mail: robk@as.arizona.edu, jpizagno@as.arizona.edu}

\author{Accepted for Publication in the ApJ}

\vskip 1.cm

\begin{abstract}

The advent of 8--10 meter class telescopes enables direct
measurement of the chemical properties in the ionized gas of
cosmologically--distant galaxies with the same nebular analysis
techniques used in local \HII\ regions.  We show that spatially
unresolved (i.e., global) emission line spectra can reliably indicate
the chemical properties of distant star-forming galaxies.   However,
standard nebular chemical abundance measurement methods (those with a
measured electron temperature from [O~III] $\lambda$4363) may be
subject to small systematic errors when the observed volume includes a
mixture of gas with diverse temperatures, ionization parameters, and
metallicities.  To characterize these systematic effects, we compare
physical conditions derived from spectroscopy of individual
\HII\ regions with results from global galaxy spectroscopy.  We
consider both low-mass, metal poor galaxies with uniform abundances and
larger galaxies with internal chemical gradients.  For low-mass
galaxies, standard chemical analyses using global spectra produce small
systematic errors in that the derived electron temperatures  are 1000 K
--- 3000 K too high due to non-uniform electron temperatures and large
variations in ionization parameter.  As a result, the oxygen abundances
derived from direct measurements of the electron temperatures are too
low, but it is possible to compensate for this effect by applying a
correction of $\Delta(O/H) = +0.1$ dex to the oxygen abundances derived
from global spectra.   For more massive metal-rich galaxies like local
spirals, direct measurements of electron temperatures are seldom
possible from global spectra.  Well-established empirical calibrations
using strong-line ratios can serve as reliable ($\pm$0.2 dex)
indicators of the overall systemic oxygen abundance even when the
signal-to-noise of the $H\beta$ and [O~III] emission lines is as low as
8:1.  We present prescriptions, directed toward high-redshift
observers, for using global emission line spectra to trace the chemical
properties of star-forming galaxies in the distant universe.

\end{abstract}


\section{Using Integrated Galaxy Spectra as Chemical Diagnostics }

Optical emission lines from \HII\ regions have long been the primary
means of gas-phase chemical diagnosis in galaxies (Aller 1942; Searle
1971; reviews by Peimbert 1975; Pagel 1986; Shields 1990; Aller 1990).
With the advent of large telescopes and sensitive spectrographs,
nebular emission lines from distant star-forming regions can probe the
chemical evolution of objects at earlier epochs (Steidel \etal\ 1996;
Kobulnicky \& Zaritsky 1998) in a complementary manner to absorption
line measurements (reviewed by Lauroesch \etal\ 1996).  Optical emission
line spectroscopy preferentially samples the warm ionized phase of the
interstellar medium in the immediate vicinity of recent star formation
events (i.e., \HII\ regions).  Madau \etal\ (1996), Lilly \etal\ (1996)
observe that the rate of star-forming is higher in the past, so
potential emission line targets are plentiful at redshifts of
$0.3<z<2$.  Compared to absorption-line studies which are limited to
lines of sight toward bright background quasars, nebular emission line
observations are possible in any galaxy with \HII\ regions.
Complications due to line width ambiguities, saturation effects,
multiple velocity components, and ionization corrections are less
severe or absent for nebular spectroscopy.  Limited signal-to-noise and
lack of spatial resolution are the two most formidable obstacles for
ground-based spectroscopic studies of cosmologically distant
\HII\ regions.  A typical ground-based resolution element of
1.0\arcsec\ corresponding to a linear size of  5.2 kpc at $z=0.5$ will
encompass entire galaxies.\footnote{For a cosmology with $H_0$=75
\kms\ Mpc$^{-1}$, $\Lambda=0$,  and $q_0$=0.1.}  Our motivation in this
paper is to explore the utility of {\it spatially-integrated} emission
line spectroscopy for studying the chemical properties of star-forming
galaxies at earlier epochs.

Osterbrock (1989) thoroughly discusses the standard techniques for
measuring the chemical properties of ionized gas.  Typically, chemical
analyses of \HII\ regions require measurement of H and He recombination
lines along with collisionally-excited lines from one or more
ionization states of heavy element species.  Oxygen is the most
commonly--used metallicity indicator in the ISM by virtue of its high
relative abundance and strong emission lines in the optical part of the
spectrum (e.g.,  [O~II] $\lambda$3727 and [O~III]
$\lambda\lambda$4959,5007).    In best-case scenarios, the electron
temperature of the ionized medium can be derived from the ratio of a
higher-excitation auroral line, such as [O~III] $\lambda$4363, to
[O~III] $\lambda$5007.  In practice, [O~III] $\lambda$4363 is difficult
to measure since it is typically only a few percent the strength of
H$\beta$ even in the most metal-poor \HII\ regions (I~Zw~18;
(O/H)=0.02(O/H)$_\odot$), and becomes unmeasurably weak in more
metal-rich environments.  The electron density of the medium may also
be constrained using the density-sensitive ratio of [S~II]
$\lambda$6713/[S~II] $\lambda$6731 or [O~II] $\lambda$3727/[O~II]
$\lambda$3729 doublets.  When bright \HII\ regions are available,
abundances of He, C, Ne, Si, S, and Ar  are well-measured in many local
galaxies.  However, in cosmologically distant \HII\ regions, only the
brightest lines may be detectable, even using the collecting area and
sensitivity of the largest telescope/instrument combinations envisioned
today.  A typical ground-based resolution element will encompass large
fractions of galaxies exhibiting a wide variety of physical
characteristics.  Are global galaxy spectra in any way indicative of
the physical properties of its ISM?

Three effects may bias spatially-integrated (global) emission line
spectra of galaxies (kpc--sized apertures) compared to the values that
would be derived from individual \HII\ regions observed with higher
spatial resolution (10---100 pc--sized apertures).

1) The aperture may include a mixture of gas or multiple \HII\ regions with 
similar metallicity but differing ionization conditions.  

2) The aperture may include a mixture of gas or multiple
\HII\ regions with significantly different metallicities.

3) The aperture may include a mixture of overlapping stellar absorption
and nebular emission features which adversely affect the measured
equivalent widths and emission line strengths.

One or several of these effects may reduce the precision of chemical
abundance determinations using global galaxy spectra.  In this paper, we
explore the magnitude of such effects by comparing chemical analyses
based upon local and global spectroscopy of nearby, well-studied
galaxies.  In Section~2 we consider the case of dwarf galaxies with
uniform gas-phase abundance distributions, but a range of ionization
conditions.  In Section~3, we consider larger spiral galaxies which
exhibit both chemical abundance and ionization variations.  In
Section~4 we consider the case of low-S/N spectra in which emission
line strengths may be affected by underlying stellar absorption.  We
also consider the possibility of measuring oxygen abundances when only
the H$\beta$ and [O~III] emission lines are detected.  The summary in
Section~5 recaps the prospects for using global galaxy spectroscopy
to study the chemical properties of star-forming objects in the early
universe.  {\it Note of Caution: } The prescriptions for deriving
gas-phase metal abundances provided here assume that the nebular
emission line spectrum is generated by photoionization from massive
stars.  Non-stellar ionizing sources, such as AGN, can  generate
emission spectra that superficially resemble photoionized
\HII\ regions.  Generally, non-stellar energy sources produce
distinctive emission line ratios compared to ordinary \HII\ regions.
Veilleux \& Osterbrock (1987) and Heckman (1980) provide diagnostic
diagrams which should be consulted to ascertain that ionizing sources
are stellar in nature before proceeding with chemical analysis.

\section{The Case of Metal-Poor Irregular and Dwarf Galaxies}

Low-mass galaxies with sizes and luminosities smaller than that of the
LMC appear to be chemically homogeneous on scales from $\sim$10 pc to 1
kpc (review by Kobulnicky 1998).  Yet, they often contain a multiplicity
of \HII\ regions, and they exhibit a variety of ionization conditions.
Localized high surface--brightness, high-ionization \HII\ regions lie
amidst a network of extended low-ionization filaments which sometimes
stretch for a kpc or more (e.g., Hunter 1994; Hunter \& Gallagher
1997).  What is the effect of mixing these  regions of disparate
ionization parameter into a single aperture?  We address this question
empirically by considering both spatially integrated and localized
measurements of ionized gas in a sample of nearby irregular galaxies.

\subsection{Data Collection}

We obtained new spectroscopic observations of the irregular and blue
compact galaxies  NGC~3125 ($\equiv$Tololo 1004-296), NGC~5253,
Henize~2-10, Tololo 35 ($\equiv$Tololo 1324-276$\equiv$IC4249) on 4-6
February 1998 using the 4 m telescope + RC spectrograph at the Cerro
Tololo InterAmerican Observatory.   The KPGL-3 grating produced a
dispersion of 1.2 \AA\ pixel$^{-1}$ across the wavelength range
3700\AA\ -- 7000 \AA\ imaged onto the 1k x 3k Loral CCD.  A slit width
of 2\arcsec\ produced a spectral resolution of 5.1 \AA.  Seeing
averaged 1-2\arcsec\ through occasional light cirrus clouds and
airmasses ranged from 1.0 to 1.6.  The spatial scale of the CCD was
0.50\arcsec\ pixel $^{-1}$. The data were reduced in the standard
fashion, by subtracting overscan bias, and dividing by a flat field
image produced from a combination of dome-illuminated exposures and
exposures of a quartz continuum lamp inside the spectrograph.  Based on
exposures of the twilight sky, we applied to the data a correction for
variations in the illumination pattern along the 5\arcmin\ slit
length.  This combination of dome and internal lamp flats was necessary
to achieve good S/N over the full wavelength range, and to correct for
scattered light within the instrument during quartz lamp exposures.
Periodic exposures of HeAr arc lamps allowed a wavelength solution
accurate to $\sim$0.5 \AA\ rms.  Multiple observations of the standard
stars GD108, GD50, Hz4, and SA95-42 (Oke 1990) with a slit width of
6\arcsec\ over a range of airmass provided a
site-dependent atmospheric extinction curve which we found to be
consistent with the mean CTIO extinction curve.  Standard star
exposures also allowed us to derive the instrumental sensitivity function
which had a pixel-to-pixel RMS of 2\%. Between five and twenty 6-pixel
apertures were defined across the nebular portion of each galaxy.  We
extracted multiple one-dimensional spectra of each object with aperture
sizes of 6 pixels which correspond to linear sizes of 20 pc to 100 pc
at the distances of the galaxies.  We used emission-free regions
20-pixels or larger on either side of the aperture to define the mean
background sky level which was subtracted from the source spectrum.

In order to measure a global emission line spectrum, we performed 5 and
10-minute drift scans of each galaxy by moving the telescope at a rate
of 1\arcsec\ $s^{-1}$ back and forth across the galaxy perpendicular to
the slit.  Summing the spatial extent of the nebular emission (100--300
pixels), we produced a global spectrum of the galaxy at the same spectral
resolution as the fixed-pointing exposures.  Figure~1 displays the
global spectrum for each galaxy in arbitrary units.  For display
purposes, we plot each spectrum a second time, multiplied by a factor
of 40.

To augment the sample, we selected four additional irregular galaxies
with both global and local nebular spectroscopy in the literature.
Kennicutt (1992---K92) presents spatially-integrated spectra for
NGC~1569, NGC~4449, and NGC~4861$\equiv$Mrk 59.  We used nebular
spectroscopy of individual \HII\ regions for these galaxies from
Kobulnicky \& Skillman (1997---NGC~1569), Talent (1980---NGC~4449), and
our own 3.5 m Calar Alto spectra (Kobulnicky, in prep---NGC~4861;
observations as described in Skillman, Bowmans \& Kobulnicky 1997).
Using the longslit data described in Kobulnicky \& Skillman (1996), we
constructed a global spectrum for NGC~4214.

We analyzed the one-dimensional localized spectra and global spectra
for each galaxy using the nebular emission--line software and
procedures described in Kobulnicky \& Skillman (1996).  The flux in
each emission line was measured with single Gaussian fits.  In the case
of blended lines or lines with low-S/N, we fixed the width and position
of the fit using the width and position of other strong lines nearby.
Logarithmic extinction parameters, c(H$\beta$), underlying stellar
hydrogen absorption, EW(abs), and the electron temperatures were
derived iteratively and self-consistently for each object using
observed Balmer line ratios compared to theoretical case B
recombination ratios (Hummer \& Storey 1987).  Table~1 lists the
emission line strengths dereddened relative to H$\beta$, c(H$\beta$),
and an adopted value for the underlying stellar hydrogen absorption,
EW(abs), for all 8 global spectra.  For the new observations obtained
here, tabulated uncertainties on the emission line strengths and
derived physical properties take into account errors due to photon
noise, detector noise, sky subtraction, flux calibration, and
dereddening.  For global spectra described in Kennicutt (1992), we
determined uncertainties empirically from the RMS in the continuum
portion of the spectrum adjacent to each line.

Analyses of the physical conditions in each galaxy follow Kobulnicky \&
Skillman (1996).  For each spectrum (except He~2-10 and NGC~4449 where
[O~III] $\lambda$4363 is not detected), we made a direct determination
of the electron temperatures, $T_e(O~III)$, from the [O~III]
$\lambda$4363 and [O~III] $\lambda$5007 lines.  We determined the
electron temperature of the lower ionization zones using the empirical
fit to photoionization models from Pagel \etal\ (1992) and Skillman \&
Kennicutt (1993),

\begin{equation} T_e(O^+)=2(T_e^{-1}(O^{++}) + 8\times10^{-5})^{-1}.
\end{equation}

\noindent The final O/H ratio involves the assumption that

\begin{equation}
{ O\over H} = {{O^{+}}\over{H^{+}}} + {{O^{++}}\over{H^{+}}}.   
\end{equation}

\noindent Analysis of the [O~I] $\lambda$6300 lines in the global and
localized spectra showed that neutral oxygen accounts for less than 4\%
of the total in all objects.  Inclusion of the O$^{0+}$ contribution
would raise the oxygen abundance by $<0.02$ dex, and is probably not
a significant factor since $O^{0+}$ co-exists
with $H^{0+}$ in photoionization models.    Nebular He~II
$\lambda$4686 was not detected in any of the global spectra except
NGC~1569, so we assume that the contribution from highly ionized
species like $O^{3+}$ is negligible.  The nitrogen to oxygen ratio, N/O
is computed assuming

\begin{equation}
{ N\over O} = {{N^{+}}\over{O^{+}}}.
\end{equation}

Equation~3 appears justified in low metallicity environments where
photoionization models for a range of temperatures and ionization
parameters indicate uncertainties of less than 20\% through this
approximation (Garnett 1990).  Table~2 summarizes the derived electron
temperatures, oxygen abundances, and nitrogen abundances for each
galaxy.  All objects except Henize 2-10 lie in the range
7.9$<$12+log(O/H)$<$8.4, typical of nearby irregular and blue compact
galaxies with metallicities between 1/10 and 1/3 the solar value.  We
use these results in the next section to investigate the potential for
measuring galaxy metallicities from global spectra.  The lack of an
[O~III] $\lambda$4363 detection in He~2-10, even in the localized
smaller apertures ($T_e<$8400 K), places this galaxy in the metal-rich
regime.  The empirical strong-line relation of Zaritsky, Kennicutt, \&
Huchra (1994) indicates 12+log(O/H)$\simeq$8.93 with probable
uncertainties of 0.15 dex.  Due to the lack of measured [O~III]
$\lambda$4363,  we exclude He~2-10 from further analyses.

\subsection{Comparison of Local and Global Spectroscopy}

Figure 2 shows the derived electron temperature for each spectrum,
along with the signal-to-noise ratio of the [O~III] $\lambda$4363
line.  Different symbols distinguish each galaxy.  Small symbols denote
measurements of  individual \HII\ regions through small apertures while
large symbols with error bars represent the global spectrum.  Figure~3
shows the resulting oxygen abundance, 12+log(O/H) for each object,
versus signal-to-noise ratio of [O~III] $\lambda$4363.  This
combination of figures reveals that apertures (within a given galaxy)
having the highest electron temperatures exhibit the lowest oxygen
abundances.    At constant oxygen abundance such a correlation is
expected since $T_e$ and the emissivity of a collisionally excited
line, $\epsilon$,  are correlated inversely as (cf., Osterbrock 1989),

\begin{equation}
 \epsilon \propto\ T_e^{-1/2} e^{h\nu/kT_e}. 
\end{equation}

Figures~2 and 3 show that the global spectra consistently indicate
higher (lower) electron temperatures (oxygen abundances) than the
individual \HII\ regions observed using smaller apertures within the
same galaxy.  In most cases, the global spectrum produces electron
temperatures (oxygen abundances) consistent with the  highest (lowest)
value derived from the smaller apertures.   One possible cause for such
a trend is that the [O~III] $\lambda$4363  line measured in the global
spectra has a lower signal-to-noise ratio, and is systematically
overestimated in the presence of significant noise, since Gaussian fits
to emission lines with very low S/N ratios are systematically biased
toward larger values.  However, since the S/N of  [O~III]
$\lambda$4363  in all the global spectra exceeds 10 (except NGC~4449)
this effect is not likely to be the major cause of a systematic
temperature deviation.  A more likely possibility is that the regions
of highest nebular surface brightness, which also tend to have the
highest electron temperatures, dominate the global spectrum.  Since the
global spectrum is the same as an intensity-weighted average, it
preferentially selects the regions of highest surface brightness which
are likely to be ionized by the youngest, hottest stars, and thus have
the highest electron temperature.  The exponential dependence of $T_e$
on the [O~III] $\lambda$4363 line biases the measured electron
temperature toward higher values.   Thus, temperature fluctuations
within the observed aperture will give rise to artificially--elevated
electron temperatures derived from collisionally-excited lines, as
observed in individual \HII\ regions and discussed extensively in the
literature (Peimbert 1967;  Kingdon \& Ferland 1995; Peimbert 1996).

Figure~4 illustrates the range of oxygen abundances derived for each
galaxy, along with the global value plotted using a large symbol and
error bars showing the uncertainties due to statistical observational
errors.  Figure~4 demonstrates more clearly that the oxygen abundances
derived from global spectra systematically lie 0.05---0.2 dex below the
median values computed from smaller apertures.  For the localized
measurements within a given galaxy, there is a strong correlation
between O/H and electron temperature as shown in Figure~5.  The slope
of this correlation within each galaxy is consistent with the
correlation expected in the presence of random electron temperature
uncertainties (solid line).  Random or systematic errors in the adopted
electron temperature can produce spurious variations in O/H that mimic
real oxygen abundance fluctuations.  The solid line in Figure~5
illustrates the direction along which the derived oxygen abundances
will deviate in the presence of significant errors on the electron
temperature.  The data are consistent with constant metal abundance
throughout each galaxy and a dispersion of $\Delta(O/H)\sim0.1$---$0.2$
dex caused by variations of $\Delta{T_e}\sim$1000---2000 K in the
derived electron temperature.  Such low-mass galaxies typically have
O/H dispersions of 0.1 dex but no measurable chemical gradient (cf.,
review by Kobulnicky 1998), even in the diffuse ionized gas at large
radii (Martin 1997).     The data in Figure~4 is consistent with the
known correlation between blue magnitude and oxygen abundance obeyed by
nearly all known galaxy types (e.g., Lequeux \etal\ 1979; French 1980;
Faber 1973; Brodie \& Huchra 1991; Zaritsky, Kennicutt, \& Huchra--ZKH;
Richer \& McCall 1995).  The luminosity-metallicity relation derived by
Skillman, Kennicutt, \& Hodge (SKH: 1989) for irregular galaxies
appears as a dashed line.  NGC~4861 and Tololo 35 deviate the most from
this trend, suggesting that either their luminosities are not well
measured, or that they are slightly over-luminous for their
metallicities compared to the majority of galaxies.  However, their
deviation is consistent with the observed dispersion in the
luminosity-metallicity relation, 0.3 dex at a fixed luminosity.

\subsection{The Effects of Variable Temperature and Ionization}

Can global spectra reliably probe the chemical content of the low-mass
galaxies?  Several effects may cause the global emission line ratios
to produce oxygen abundance estimates $0.05$---$0.2$ dex lower than
those derived from small aperture observations of individual
\HII\ regions.  Since the abundances derived from collisionally-excited
lines are sensitive to the assumed electron temperature, the most
likely cause involves temperature uncertainties or temperature
fluctuations within the observed region.  Peimbert (1967) first
discussed the effects of temperature fluctuations within
\HII\ regions.  Since then a variety of authors have investigated their
effects on the measured chemical composition of the Orion nebula
(Walter, Dufour, \& Hester 1992), planetary nebulae (Liu \& Danziger
1993; Kingdon \& Ferland 1995), on primordial helium abundance
measurements (Steigman, Viegas, \& Gruenwald 1997), and in evolving
starbursts (P\'erez 1997).   Typically, the magnitude of the
temperature fluctuations is described by Peimbert's (1967) parameter,
$t^2$, the root mean square of the temperature variation.
Observational constraints suggest that $t^2$ ranges from $t^2=0.03$ for
most planetary nebula to $t^2=0.1$ for a few planetary nebulae (Liu \&
Danziger 1993) and giant extragalactic \HII\ regions like NGC~2363
(Gonzalez-Delgado \etal\ 1994).   In addition to temperature
fluctuations within individual star-forming regions, many galaxies
exhibit temperature gradients of 1000 K -- 3000 K (NGC~5253---Walsh \&
Roy 1989; II~Zw~40; Walsh \& Roy 1993; I~Zw~18---Martin 1996;
NGC~4214---Kobulnicky \& Skillman 1996; NGC~1569---Kobulnicky \&
Skillman 1997).  Thus, real galaxies containing one or more giant
\HII\ regions will contain an arbitrary mixture of gas with differing
electron temperatures.  They will produce global spectra that cannot
easily be characterized by the simple single-temperature, two-zone
approximation ($T_e(O^{+})$ and $T_e(O^{++})$ commonly used.  Peimbert
(1967) and the subsequent researchers have shown that in the presence
of temperature fluctuations, collisionally-excited lines indicate
electron temperatures which are 1000 K -- 4000 K higher than $T_e$
measurements from recombination lines.

A second factor that may influence chemical determinations from global
galaxy spectra, even in the absence of temperature fluctuations, is
variation of the ionization parameter, $U=Q_{Ly}/(4\pi R^2~n_H~c)$, the
number density of ionizing photons.  Diffuse, inter-\HII\ region gas
(i.e., Diffuse Ionized Gas; DIG) with low ionization parameter
($\log{U}\simeq-3.5$) accounts for 20\% to 50\% of the Balmer line
emission in irregular and spiral galaxies (Hunter \& Gallagher 1990;
Martin 1997; Ferguson \etal\ 1996).  Diffuse ionized gas appears to be
mostly photo-ionized, perhaps with a shock-excited component (Hunter \&
Gallagher 1990; Martin 1997).  Diffuse ionized gas in irregular
galaxies exhibits higher [O~II]/[O~III] ratios ($1<[O~II]/[O~III]<10$)
than \HII\ regions  ($0.1<[O~II]/[O~III]<2$).  This is a signature of a
decreasing ionization parameter, as the distance from the ionizing O
and B stars increases (Hunter \& Gallagher 1990; Hunter 1994; Martin
1997).  The electron temperature of this diffuse gas is not well
constrained, however.  For the six galaxies with local and global
spectroscopy, there  is weak evidence for a correlation between the
ionization parameter, as measured by $O^+/O^{++}$ and the magnitude of
the offset between the global-derived oxygen abundance, and the mean
O/H derived from localized spectra.  Figure~6 shows the global
$O^+/O^{++}$ for each galaxy versus the difference between the mean
(O/H) for individual \HII\ region measurements and the (O/H) ratio
derived from the global spectra for each object.  The data have a
linear correlation coefficient of 0.52, consistent with the existence
of a correlation at the 70\% confidence level.  The largest deviations
are nearly $-0.15$ dex for NGC~1569 and NGC~4861 which have the lowest
$O^+/O^{++}$ values of 0.2, consistent with the smallest contribution
from diffuse ionized gas with a low ionization parameter.  None of the
objects in this sample appear to be dominated by low-ionization gas.
In all cases, $O^+/O^{++}>1.0$, so we cannot address empirically the
impact of large contributions from diffuse ionized gas with these
data.  Global and spatially-resolved observations of large irregular
galaxies dominated by diffuse gas (e.g., NGC~1800, NGC~3077, NGC~4449)
can address this issue.

To simulate the possible effects of temperature fluctuations and
varying ionization parameter on chemical analysis, we construct a set
of six emission line spectra which characterize interstellar gas with a
realistic range of temperatures and ionization parameters, but with a
common oxygen abundance, 12+log(O/H)=8.0.  We refer to these six
spectra as ``basis spectra'' which we mix in different proportions, to
simulate the effects of inhomogeneous temperature and ionization
conditions on chemical analysis.  Table~3 summarizes the physical
parameters of the basis spectra.  The ``standard'' spectrum, $S$, with
an electron temperature, $T_e(O^{++})$, of 13,000 K, and
$O^+/O^{++}=0.2$, represents a relatively high ionization parameter of
$\log{U}\simeq-3$ in the models of  Stasi\'nska (1990).  Spectra
T,U,V,W,X have $T_e(O^{++})$ of either 13,000 K, 11,000 K, or 9000 K.
The temperature differences between spectra, $\Delta{T_e}(O^{++})$=2000
K and $\Delta{T_e}(O^{++})$=4000 K correspond roughly to $t^2=0.08$ and
$t^2=0.14$ in the notation of Peimbert (1967).  The former is within
the $t^2$ values observed in actual nebulae, while the latter
represents an upper bound to observed values.  Approximate ionization
parameters range from $\log{U}\simeq-3$ or $\log{U}\simeq-4$,
parameterized here as $O^+/O^{++}=0.2$ or $O^+/O^{++}=2.0$.  This range
reflects the variations observed in irregular galaxy \HII\ regions
(Martin 1997) and will serve for the modeling required here.  However,
the $O^+/O^{++}$ ratio in diffuse ionized gas can reach as high as
$\sim$7 for line ratios [O~II]/[O~III]$\simeq$10 (Martin 1997).

We begin by mixing the emission line spectrum of the standard sample,
$S$, with the spectra of samples T,U,V,W,X in ratios of 80:20, 50:50,
20:80, and 10:90.  Our standard nebular software is then
used to analyze the resulting composite spectrum.  Figure~7 displays
the resulting electron temperatures, oxygen abundances, and
$O^+/O^{++}$ ratios derived for the composite samples.  Different
symbols denote samples comprised of 80\% (filled squares), 50\% (filled
circles), 20\% (open circles), and 10\% (open squares) of the standard
sample, $S$.  Crosses designate the physical conditions of the basis
spectra, T,U,V,W,X, and S, used to construct the composite spectra.
Line styles denote combinations of spectra T+S (dash-dot-dot line), U+S
(solid line), V+S (dashed line), W+S (dotted line) and X+S (dash-dot
line).

Figure~7 (top panel) shows that, as the spectrum S
($T_e(O^{++})$=13,000 K, $O^+/O^{++}=0.2$) is increasingly diluted
with gas from spectrum U ($T_e$=11,000 K, $O^+/O^{++}=0.2$), the
measured electron temperature decreases smoothly, while the measured
oxygen abundance becomes slightly underestimated by up to 0.02 dex.  For
example, when the mixture of the composite spectrum is 80\% S and 20\%
U (filled square on the solid line) the derived electron temperature is
12,750 K.  The derived oxygen abundance is 12+log(O/H)=7.99, an
underestimate by 0.01 dex.  As the composite spectrum becomes 50\% S
and 50\% U (solid line, filled circle) the derived oxygen abundance is
underestimated by 0.02 dex.  As the composite mixture becomes dominated
by spectrum U (open circle and then open square) the derived physical
conditions converge once again toward the temperature of basis spectrum
U (11,000 K) and toward the initial oxygen abundance of both spectra,
12+log(O/H)=8.0.

When spectrum S is combined with spectrum
V ($T_e$=11,000 K, $O^+/O^{++}=0.2$) the measured deviation from
12+log(O/H)=8.0 becomes more pronounced, up to 0.05 dex.  The lower
panel of Figure~7 shows a smooth progression in the measured
$O^+/O^{++}$ ratio from 0.2 to 2.0.  The mixture of
spectrum S with spectrum W ($T_e$=9000 K,
$O^+/O^{++}=0.2$) shows a substantial systematic deviation from
constant metallicity.  When the fraction of the lower temperature,
lower ionization gas represented by spectrum W reaches 50\% to 90\% of
the total emission line flux, the oxygen abundance is underestimated by
0.1 to 0.2 dex!  A similar underestimate of the oxygen abundance
results from spectrum X ($T_e$=9000 K, $O^+/O^{++}=2.0$).   Figure~7
demonstrates that temperature fluctuations, modeled in the simplest
possible way as a two-temperature medium, are the primary cause for
over-estimation of the electron temperature and under-estimation of the
oxygen abundance.  Ionization parameter variations further exacerbate
the systematic underestimate of oxygen abundances.  Figure~7 shows that
the addition of even a small quantity of high temperature gas (10\%)
creates a significant overestimate of the mean electron temperature.

Figure~7 demonstrates that the measured electron temperature,
$T_e(O^{++})$, is sensitive to the physical conditions in the hottest
$O^{++}$ zone of an \HII\ region due to the strongly non-linear
dependence of [O~III] $\lambda$4363 on $T_e$.  Furthermore, the
electron temperature, $T_e(O^{++})$, in the $O^{++}$ zone is usually
different than the electron temperature in lower-ionization $O^+$,
$N^+$, $S^+$ zones at larger radii.   Since $T_e(O^{+})$ or
$T_e([S~II])$ is seldom measured directly, most nebular analyses
procedures, including ours, adopt an empirical estimate based on the
measured $T_e(O^{++})$ and photo-ionization modeling (Pagel
\etal\ 1992; Skillman \& Kennicutt 1993).  However, in the presence of
temperature fluctuations, or, in the limiting case of a two-temperature
medium, the derived [O~III] electron temperatures are weighted toward
the high-temperature medium.  Application of a naively--computed
$T_e(O^{++})$ artificially decreases the total abundance by
underestimating the oxygen abundance of both the  $O^{++}$ zone and
the  $O^{+}$ zones.

A additional underestimate of the total metal abundance results if a
large fraction of the nebular medium has a low ionization parameter.
This is because estimates of $T_e(O^{+})$ derived from $T_e(O^{++})$
via an empirical relation based on photoionization models (Equation~1)
will be inappropriately high.  They systematically overestimate
$T_e(O^{+})$, and underestimate the oxygen contribution from the
dominant $O^+$ zones.  While photoionization models can simulate the
expected ionization structure of an ideal Stromgren sphere under a
variety of ionization conditions, they do not take into account
emission from the extended ionized (predominantly low-U) filaments and
shells that are seen in actual galaxies.    Since important factors
such as the porosity of \HII\ regions, the ionizing source of the
extended shells and filaments, and the origin of the variation in DIG
content from galaxy to galaxy are still uncertain, it is unlikely that
these structures will soon be incorporated into photoionization
models.

\section{The Case of Spiral Galaxies}

Like their low-mass counterparts, star-forming spiral galaxies also
exhibit a range of ionization and temperature conditions throughout
their ISM.  However, except those with strong optical bars, they also
show considerable radial chemical gradients, often exceeding an order
of magnitude (Searle 1971; Villa-Costas \& Edmunds 1992;  Zaritsky,
Kennicutt \& Huchra 1994; Martin \& Roy 1994).  Global spectra of
spiral galaxies will necessarily encompass a wide range of
metallicities.  In this section we consider whether it is possible to
use global galaxy spectra, even in the presence of true chemical
variations, to measure a ``mean'' or ``indicative'' systemic
metallicity.

In high-metallicity \HII\ regions ($12+{\log}(O/H)\geq8.5$), the
temperature sensitive [O~III] $\lambda$4363 line is very weak, and it
is seldom detected.  Nevertheless, the oxygen abundance can be
estimated using only the [O~II] $\lambda$3727, [O~III]
$\lambda\lambda$4959,5007, and H$\beta$ lines using the method proposed
by Pagel \etal\ (1979) and subsequently developed by many authors.  For
high-metallicity  ($12+{\log}(O/H)\geq8.5$) \HII\ regions, there exists
a monotonic relationship between the ratio of observed
collisionally-excited emission line intensities,

\begin{equation}
R_{23}\equiv(I_{3727}+I_{4959}+I_{5007})/H\beta, 
\end{equation}

\noindent and the oxygen
abundance of the nebula.  In practice, only one of the [O~III] lines is
required, since $I_{5007}\simeq2.87I_{4959}$ for all temperatures and
densities encountered in \HII\ regions.  In the most metal-rich
\HII\ regions, $R_{23}$ is a minimum because the high metal abundance
produces efficient cooling, reducing the electron temperature and the
level of collisional excitation.  $R_{23}$ increases in progressively
more metal-poor nebula since lower metal abundance means reduced
cooling, elevated electron temperatures, and a higher degree of
collisional excitation.  However, the relation between $R_{23}$ and O/H
becomes double valued below about 12+log(O/H)=8.4 ($Z=0.3 Z_\odot$).
Figure~8 illustrates this double-valued behavior.  In Figure~8, we plot
a variety of published calibrations between $R_{23}$ and O/H, including
McGaugh (1991: solid line), Zaritsky, Kennicutt, \& Huchra, (1994:
dashed line), McCall, Rybski, \& Shields (1985: dotted line), Edmunds
\& Pagel (1984: dash-dot line), and Dopita \& Evans (1986: dash-dot-dot
line).  On the upper, metal-rich branch of the relationship, the
various calibrations show a dispersion of 0.2 dex at a fixed value of
$R_{23}$.   This dispersion represents the inherent uncertainties in
the calibration which are based on photoionization modeling and
observed \HII\ regions (see original works for details).   For metal
abundances progressively lower than 12+log(O/H)$\simeq$8.2, $R_{23}$
decreases once again.  On this lower branch, although the reduced metal
abundance further inhibits cooling and raises the electron
temperatures, the intensities of the [O~II] and [O~III] lines drop
because of the greatly reduced abundance of oxygen in the ISM.
 On the lower, metal-poor branch of the relationship, a second
parameter, the ionization parameter $U$, becomes important, in
additional to $R_{23}$.  This can be seen in the offset between the
three solid lines from the calibration of McGaugh (1991).  A 
varying ionization parameter may lead to a similar value of
$R_{23}$ for different oxygen abundances.  In Figure~8 we represent
the approximate ionization parameter in terms of the easily observable
line ratio [O~III]/[O~II].    The solid lines show
the oxygen abundance as a function of $R_{23}$ for
[O~III]/[O~II] = 10, 1.0, and 0.1 which correspond
(very roughly) to ionization parameters, $U$, of $10^{-1}$, $10^{-2}$,
and $10^{-4}$. 

Figure~8 serves as a useful diagnostic diagram for finding the oxygen
abundances of nebulae when the electron temperature is not measured
directly.    The typical uncertainties using this empirical oxygen
abundance calibration are $\pm0.15$ dex, but are larger, ($\pm$0.25
dex) in the turn-around region near 12+log(O/H)$\sim$8.4 when
${\log}(R_{23})>0.7$.  The most significant uncertainty involves
deciding whether an observed object lies on the upper, metal-rich
branch of the curve, or on the lower, metal-poor branch of the curve.
For instance, a measurement of $\log(R_{23})=0.0$ could indicate either
an oxygen abundance of 12+log(O/H)$\simeq$7.2 or
12+log(O/H)$\simeq$9.1.  Knowledge of either the luminosity of the
galaxy or the [N~II] $\lambda$6584 intensity can help break the
degeneracy.  Because star-forming galaxies of all types {\it in the
local universe} follow a luminosity-metallicity correlation (e.g.,
Lequeux \etal\ 1979; Talent 1980; Skillman, Kennicutt, \& Hodge 1989,
ZKH), objects more luminous than $M_B\simeq-18$ have metallicities
larger than 12+log(O/H)$\simeq$8.3, placing them on the upper branch of
the curve.  However, it has not yet been established whether galaxies
at earlier epochs conform to the same relationship as local galaxies.
An even better discriminator is the ratio [O~III]$\lambda$5007/[N~II]
$\lambda$6584 which is usually greater than $\sim$100 for galaxies with
12+log(O/H)$>$8.3 on the metal-rich branch (Edmunds \& Pagel 1984).
This is because objects which are considerably enriched in oxygen are
generally more nitrogen-rich as well, while the most metal-poor
galaxies on the lower branch of the $R_{23}$ relation have very weak
[N~II] lines.  Figure~A1(b) of Edmunds \& Pagel (1984) shows the
monotonic sequence of the ratio [O~III] $\lambda$5007/[N~II]
$\lambda6584$ as a function of metallicity.  If the [N~II]
$\lambda6584$ line can be measured, we believe it will provide the most
useful way to break the $R_{23}$ degeneracy in the absence of a
measured electron temperature.

\subsection{Individual \HII\ Regions and Global Spiral Spectra}

The compilation of \HII\ region spectra in spiral galaxies
presented by Zaritsky, Kennicutt, \& Huchra (1994; ZKH) provides
an excellent dataset to explore the utility of global metallicity
measurements in galaxies with chemical gradients. 
We compiled a subset of spectra for 194
HII regions in 22 galaxies from ZKH and Kennicutt \& Garnett (1996).
We use our own remeasurements of the emission line
ratios in the subsequent analysis.  

We estimate the integrated [O~II]/\hbeta\ and [O~III]/\hbeta\ emission
line ratios for each galaxy in the following way.  Spectra of the
individual HII regions from ZKH are used to define the range of
[O~II]/\hbeta\ and [O~III]/\hbeta\ as a function of galactocentric
radius.  For NGC~5457 (M101) we use data from Kennicutt \& Garnett
(1996).  In order to provide a meaningful estimate, we restrict the
analysis to galaxies with at least 8 measured HII regions, spanning
most of the radial range over which significant star formation takes
place, and for which data on the radial distribution of
\halpha\ emission are available.  The actual number of HII regions
measured ranges from 8 in NGC~4725 and NGC~5033 to 42 in M~101.
Typical values lie in the range 10--20.  We subdivide each disk into
5--13 equal-sized radial zones, with the number of zones depending on
the number and distribution of HII regions; the mean line ratios in
each zone are derived from the averages of the individual nebular
line ratios.  In a few instances a zone did not contain a measured HII
region, and in such cases the local line ratios were interpolated from
the two adjacent zones.

In order to derive disk-integrated line ratios, we compute a weighted
radial average with the spectrum at each radius weighted by the
relative \halpha\ surface brightness at each radius and the area
contained in each zone.  The \halpha\ radial profiles are derived from
\halpha\ CCD images from Martin \& Kennicutt (1998) and unpublished
imaging from the original ZKH program, using an ellipse-fitting surface
photometry routine.  We sum [O~II]/\hbeta\ and [O~III]/\hbeta\ ratios
for each galaxy to derive an integrated $R_{23}$ index for each galaxy,
as given in Equation~5.  In the calculations presented 
here we use the reddening-corrected HII region line strengths,
convolved with the observed (uncorrected) H-alpha emission distributions,
because this most faithfully duplicates the
actual weighting when the integrated spectrum of a galaxy is
observed.   Table~4 lists the integrated line strengths and $R_{23}$
values for each galaxy, along with the number of HII regions used to
derive these average values.

In Figure~9 we plot the integrated [O~II] (log [O~II]/\hbeta) versus
[O~III] (log [O~III]/\hbeta) line strengths for each galaxy determined
in this way (large triangles), as well as the [OII] and [O~III]
strengths for the individual HII regions in the same galaxy sample
(dots).  This comparison shows that the most of the integrated spectra
lie along the same excitation/abundance sequence that is defined by the
individual HII region.  Such a correspondence would not necessarily be
predicted {\it a priori}, because the HII region abundance sequence is
not entirely linear, and one might expect the average spectra to be
systematically displaced from the sequence.  The tendency for the
integrated spectra to lie on the excitation sequence partly reflects
the limited range of abundance and excitation in many disks (especially
barred systems), and the radial concentration of star formation in
others.

It is clear from Figure 9 that the integrated [O~II] and [O~III] line
strengths mimic those of individual HII regions, but how do the
corresponding ``mean" abundances compare to the actual abundances in
the disks?  In order to address this question we apply the $R_{23}$
calibration of ZKH to the integrated $R_{23}$ values in Table 4, to
estimate a mean abundance for each galaxy.  The ZKH calibration is only
valid for HII regions which lie on the metal-rich branch of the R23
relation (see discussion above in Section 3), but this condition is
satisfied for the sample considered here, with the exception of the
outermost HII regions in M101, where $T_e$-based abundances were used.

Figure 10 illustrates how well the weighted-average mean nebular
spectrum characterizes the overall galaxy abundance.  In Figure 10 we
compare the empirical abundances derived from the integrated spectra
with the actual disk abundances from ZKH, measured at 0.4 times the
corrected isophotal radius (25.0 mag arcsec$^{-2}$ in $B$).  The latter
abundances are listed in Table 4, along with two other characteristic
abundances from ZKH, that measured at a fixed linear radius of 3 kpc
(using distances given in ZKH), and the abundance at 0.8 exponential
scale lengths (also given in ZKH).  Figure 10 shows that the integrated
[O~II] and [O~III] emission-line strengths provide an excellent
estimate of the mean abundance of the disk, corresponding roughly to
the value at 40\%\ of the optical radius of the disk (indicated for
reference by the solid line in Figure 10).  The dispersion about the
mean relation is very small, $\pm$0.05 dex.  On the basis of these
results we conclude that the beam-smearing effects from sampling large
numbers of HII regions in the disk have a small effect on
characterizing the mean abundance of galaxies, even those with
substantial abundance gradients, relative to the intrinsic
uncertainties in the R23 method itself.

Although these results offer assurances that the effects of averaging
the composite spectra of large numbers of discrete HII regions do not
seriously hamper the measurement of a mean disk abundance, there are
other systematic effects, not incorporated into our simulations, that
may affect the derived abundances more significantly.  One such effect
is dust reddening, which will tend to depress the flux of [O~II]
$\lambda$3727 relative to [O~III] and \hbeta\ and tend to cause the
mean abundance to be overestimated (for galaxies on the upper branch of
the $R_{23}$--O/H relation).  In our simulations we used the measured
(i.e., reddened) line strengths when simulating the integrated spectra,
so this effect is already incorporated into the comparisons shown in
these Figures.  To test the effects of reddening on the integrated
spectra we carried out an identical comparison using
reddening-corrected spectra, and the resulting O/H abundances change
only slightly, increasing by 0.06 dex on average.  A potentially more
serious effect may be the contribution to the integrated spectrum of
diffuse ionized gas, as discussed in Section~2.3.  This gas tends to be
characterized by stronger [O~II] emission and weaker [OIII] emission
than discrete HII regions of the same abundance (e.g., Hunter 1994,
Martin 1997).  However the same data show that the sum of [O~II] and
[O~III] line strengths ($R_{23}$) does not change substantially, so
including the diffuse gas may preserve the information on mean
abundances, at least in the metallicity range probed here.  However it
would be useful to test this conclusion directly using actual
integrated spectra of spiral galaxies.  Finally, the effects of stellar
absorption of \hbeta\ in integrated spectra of spirals can be very
significant (see Section 4.2) and must be corrected for in the measurements of the
integrated spectrum (Kennicutt 1992).  Overall, the practicalities of
measuring integrated emission-line strengths in spiral galaxies
probably pose a much more challenging problem than the actual
interpretation of the composite nebular spectrum.

\section{The Effect of Underlying Stellar Absorption and Low Signal-to-Noise
Spectra}

In the previous two sections we show that global galaxy spectra can
provide reliable information on the chemical properties of distant
galaxies which exhibit high equivalent width emission lines.  However,
except for the most vigorously star-forming objects, the integrated
spectra of most large galaxies are dominated by stellar continuum,
rather than nebular emission.  In some galaxies with strong
post-starburst stellar populations, the nebular Balmer line emission is
erased by stellar atmospheric absorption.  Only 7 of the 24 global
spectra of normal spiral and elliptical galaxies obtained by Kennicutt
(1992) show measurable [O~II], [O~III], and H$\beta$ emission lines
needed for empirical oxygen abundance determinations.  However, a
larger fraction of irregular and peculiar spiral galaxies show the
requisite emission lines.  In this section, we discuss the effects of
low signal-to-noise spectra and stellar absorption which limit the
precision of nebular abundance measurements from global spectra.

\subsection{Uncertainties due to Low Signal-to-Noise}

Because the empirical calibration between $R_{23}$ and oxygen abundance
has an intrinsic uncertainty of 0.2 dex (40\%), the total error budget
will be dominated by this uncertainty even when the observed emission
line spectra have low signal-to-noise.  For example, a spectrum with a
signal-to-noise of 8:1 (12\%) on each of the [O~II], [O~III], and
H$\beta$ emission lines will yield an uncertainty of 25\% on $R_{23}$
or $\delta({\log}R_{23})\simeq$0.1  This observational uncertainty
propagates into an O/H uncertainty of 0.1--0.2 dex, depending on the
local slope of the calibration curve in Figure~8.  This quantity is
smaller than, or comparable to the uncertainty of the calibration
curve, $\sim$0.2 dex, based on photoionization modeling.  Thus, to
within the accuracy of the strong-line calibration, even modest
signal-to-noise spectra can yield useful indications of a galaxy's
gas-phase oxygen abundance.

\subsection{Uncertainties due to Stellar Balmer Line Absorption}

In high signal-to-noise nebular spectra, the observed ratios of
H$\alpha$/H$\beta$, H$\gamma$/H$\beta$, and H$\delta$/H$\beta$,
compared to theoretical values, simultaneously constrain the amount of
reddening from intervening dust, and the degree to which the stellar
atmospheric absorption lines reduce the measured nebular Balmer
emission.\footnote{This assumes that the underlying EW of the Balmer
absorption is the same for the strongest Balmer lines.  This assumption
appears to be approximately valid for most star-forming populations
(Olofsson 1995).} In practice, global galaxy spectra, especially at
high redshift, will seldom have the signal-to-noise and wavelength
coverage necessary to decouple these effects.  In the absence of
high-quality Balmer line measurements, we recommend that a statistical
correction be applied to the measured strength of the Balmer emission
lines, principally H$\beta$.  For integrated galaxy spectra presented
here, and for the global spectra presented in Kennicutt (1992) the
amount of underlying stellar absorption, Abs(H$\beta$), runs from
1\AA\ to 6 \AA, with a mean of 3 \AA.  (see also McCall, Rybski, \&
Shields 1985; Izotov, Thuan, \& Lipovetsky 1994 for a sample of
observations).  We recommend that a statistical correction of $+3\pm2$
\AA\ be applied to H$\beta$ measurements from global galaxy spectra.
When the signal-to-noise ratio of H$\beta$ is low, the uncertainty of
$\pm$2 \AA\ will act as an additional error term.  For example, for
spectra where H$\beta$ has an equivalent width of 8, the statistically
important $+3\pm2$ \AA\ correction on EW(H$\beta$) introduces an
additional 25\% uncertainty on the strength of H$\beta$.  This
uncertainty must be accounted for in the total error budget.

\subsection{Uncertainty due to Un-measured Ionization Species}

An empirical measurement of the oxygen abundance using the method
outlined in Section~3 requires measurements of the emission line
strengths for both of the dominant species of oxygen, [O~III]
$\lambda\lambda$4959,5007, [O~II] $\lambda$3727, and H$\beta$.
However, complications due to limited wavelength coverage, or
contamination from night sky lines and atmospheric absorption bands may
preclude measurement of the necessary emission lines for objects at
unfavorable redshifts.  Fortunately, measurement of either [O~III]
$\lambda$4959 or [O~III] $\lambda$4959, along with [O~II] $\lambda$3727
and H$\beta$, is sufficient to measure oxygen abundances.   The fixed
theoretical ratio of $\lambda$5007/$\lambda$4959 is $\sim$2.9 for all
electron temperatures and densities encountered in photoionized
\HII\ regions (although ratios as higher than 3 are sometimes measured
in \HII\ regions, and are more commonly seen in supernova remnants).  
Thus, the strength of one line can be computed from the other, and the
accuracy of the O/H determination is not diminished.

Oxygen abundances become highly uncertain in the case where [O~II]
$\lambda$3727 is not measured.   [O~II] $\lambda$3727 is a crucial
diagnostic for indicating the ionization parameter, fraction of diffuse
ionized gas, and possible contamination from AGN-like excitation
mechanisms.  $O^+$ may be the dominant ionization state in
low-ionization \HII\ regions.  The global spectra in Kennicutt (1992)
indicate that $O^+$ is the dominant ionization species in more than
half of the 24 galaxies with suitable emission lines.  Ratios of
$O^+/O^{++}$ range from 0.10 to 4.0.  Neglecting the contribution from
$O^+$ may result in errors as small as 10\% in the case of
high-ionization nebulae, up to a factor of 4 in galaxies dominated by
low-ionization gas.    However, the metallicity-excitation sequence
observed in Figure~9 between [O~III]/H$\beta$ and [O~II]/H$\beta$
suggests that, {\it for the majority of normal galaxies on the
metal-rich branch of the $R_{23}$ relation}, a measurement of
[O~III]/H$\beta$ can constrain the value of [O~II]/H$\beta$ to within a
factor of 3.  The converse is not true.  An observed ratio
[O~II]/H$\beta$ does not correspond to a any well-constrained
[O~III]/H$\beta$ ratio since the scatter in [O~III]/H$\beta$ at a given
[O~II]/H$\beta$ exceeds a factor of 10.  Attempts to estimate oxygen
abundances without an [O~II]/H$\beta$ ratio must carry very large
uncertainties of 0.5 dex, while attempts to measure oxygen abundances
without an [O~III]/H$\beta$ ratio have uncertainties exceeding 1 dex.

\section{Prospects for Measuring the Metallicities of 
High-Redshift Emission Line Galaxies}

We have shown that global galaxy spectra which are dominated by normal
H~II regions can provide reliable information on the gas-phase chemical
abundances, even in objects with variable gas temperature and chemical
properties.  In summary, chemical analysis using nebular optical
emission lines falls into three regimes.

1). {\it [O~III] $\lambda$4363 is detected along with [O~II]
$\lambda$3727, [O~III] $\lambda\lambda$4959,5007 and H$\beta$}:  In the
best case scenario, the  [O~III] $\lambda$4363 line can be used to
derive an electron temperature for the emitting gas, and chemical
abundance ratios can be estimated using standard nebular analysis
techniques (e.g., Osterbrock 1989).  In local galaxies, [O~III]
$\lambda$4363 is generally detected only in galaxies with
12+log(O/H)$\leq$8.4 ($Z\leq0.3~Z_\odot$).  Based on spatially-resolved
and global spectra of local irregular galaxies which are chemically
homogeneous but contain varying temperature and ionization conditions,
we find that the  [O~III] $\lambda$4363 line strength provides a firm
upper limit on the mean electron temperature of the ionized gas.  The
oxygen abundance derived using this $T_e$ is therefore a firm lower
limit.  Our empirical results in local galaxies, combined with modeling
realistic mixtures of \HII\ regions with varying physical
conditions, suggests that a statistical correction factor of
$\Delta(O/H)=+0.1$ or $\Delta{T_e}=-1000$ K be applied to physical
parameters derived from global galaxy spectra.

2) {\it [O~III] $\lambda$4363 is not detected but [O~II] $\lambda$3727,
[O~III] $\lambda\lambda$4959,5007 and H$\beta$ are measured}:  Most of
the time, the temperature sensitive [O~III] $\lambda$4363 will not be
detected due to either limited signal-to-noise or intrinsically weak
lines in metal-rich nebulae. In this case, the empirical
strong-line calibration in Figure~8  can still be used to derive an
oxygen abundance to within $\pm$0.2 dex (the uncertainty in the model
calibrations) if the [O~III] $\lambda\lambda$5007,4959, [O~II]
$\lambda$3727, and H$\beta$ lines are measured with a signal-to-noise
of at least 8:1.  A major difficulty with this method is that the
relation between oxygen abundance and $R_{23}$ is double valued,
requiring some assumption or rough $a~priori$ knowledge of a galaxy's
metallicity in order to locate it on the appropriate branch of the
curve.  We suggest that the [N~II]/[O~III] line ratios
may be useful in breaking this degeneracy, if they can be measured.

Analytic fits to the curves in Figure~8 may assist in computing the
oxygen abundances from measured line ratios.  Zaritsky, Kennicutt, \&
Huchra (1994) provide a polynomial fit to their average of three
previous calibrations shown in Figure~8.  This mean relation is good
for the upper, metal-rich regime only:

\begin{equation}
12+log(O/H)=9.265-0.33x - 0.202x^2 - 0.207x^3 - 0.333x^4 \ (ZKH~1994)
\end{equation}

\noindent where 
\begin{equation}
x\equiv\ \log{R_{23}}\equiv\ \log\Biggr({{[O~II] \lambda3727 + [O~III]
\lambda\lambda4959,5007}\over{H\beta}}\Biggr)
\end{equation}

McGaugh (1991) computed a more extensive calibration based on a set of
photoionization models which take into account the effects of varying
ionization parameter in both the metal-rich and metal-poor regimes as
shown in Figure~8.  McGaugh (1998) provides analytic expressions for
the metal-poor (lower) branch,

\begin{equation}
12+log(O/H)_{l} = 12 -4.944+0.767x+0.602x^2-y(0.29+0.332x-0.331x^2),
\end{equation}

\noindent and for the metal-rich (upper) branch,

\begin{eqnarray}
12+log(O/H)_{u} = 12 -2.939-0.2x-0.237x^2-0.305x^3-0.0283x^4-  \cr
	y(0.0047-0.0221x-0.102x^2-0.0817x^3-0.00717x^4),
\end{eqnarray}

\noindent where

\begin{equation}
y \equiv\ \log(O_{32}) \equiv\ \log\Biggr(
	{{[O~III]\lambda\lambda4959,5007}\over{[O~II]\lambda3727}}\Biggr)
\end{equation}

These analytic expressions to the semi-empirical calibration
of McGaugh (1991, 1998) fit the models to within an RMS of $\leq$0.05 dex.

3) {\it Only [O~III] $\lambda\lambda$4959,5007 and H$\beta$ are
measured}:  Very crude oxygen abundances may be derived from the ratio
of [O~III] $\lambda\lambda$4959,5007/H$\beta$ even if the [O~III]
$\lambda$4363 and [O~II] $\lambda$3727 are not measured.  This requires
both an assumption about the objects location on the bi-valued $R_{23}$
relation, {\it and} an assumption about the ionization parameter which
leads to an estimate of the [O~II] $\lambda3727$ line strength.
Uncertainties in this case must exceed 0.5 dex.

4)  {\it The spectrum has a signal-to-noise less than 8:1, or one of
the necessary emission lines are not measured}:  In this worst-case
scenario, the uncertainties on the O/H ratio will exceed a factor of 3
(0.5 dex).  Conclusions about the chemical nature of the object under
consideration will be speculative at best.

Given that star-forming regions appear to be plentiful at higher
redshifts, the prospects appear good for measuring chemical abundances
in distant galaxies, even with coarse spatial resolution of
ground-based telescopes.  The coming generation of near-infrared
spectrographs on large telescopes will make it possible to trace the
chemical evolution of the universe using emission line regions in a
manner complementary to absorption-line techniques.

\acknowledgments    We are grateful to Mauricio Navarrete for his
expertise with observations and calibration at CTIO and Sabine M\"ohler
for assistance at Calar Alto.  We appreciate a copy of the electronic
galaxy spectra from Dennis Zaritsky, and helpful conversations with Max
Pettini, Crystal Martin, and Stacy McGaugh.  H.~A.~K. and R.~C.~K thank
the Aspen Center for Physics for the opportunity to collaborate on this
research during a three-week workshop on star formation in June 1998.
H.~A.~K appreciates hospitality at the 1998 Guillarmo Haro
International Program for Advanced Studies in Astrophysics at the
Instituto National de Astrofisica Optica y Electronica (INAOE) in
Puebla, Mexico where this work was completed.  R.~C.~K. and J.~P.  were
supported by NSF grant AST-9419150.  Support for H.~A.~K was also
provided by NASA through grant \#HF-01094.01-97A awarded by the Space
Telescope Science Institute which is operated by the Association of
Universities for Research in Astronomy, Inc. for NASA under contract
NAS 5-26555.

\begin{figure}
\centerline{\psfig{file=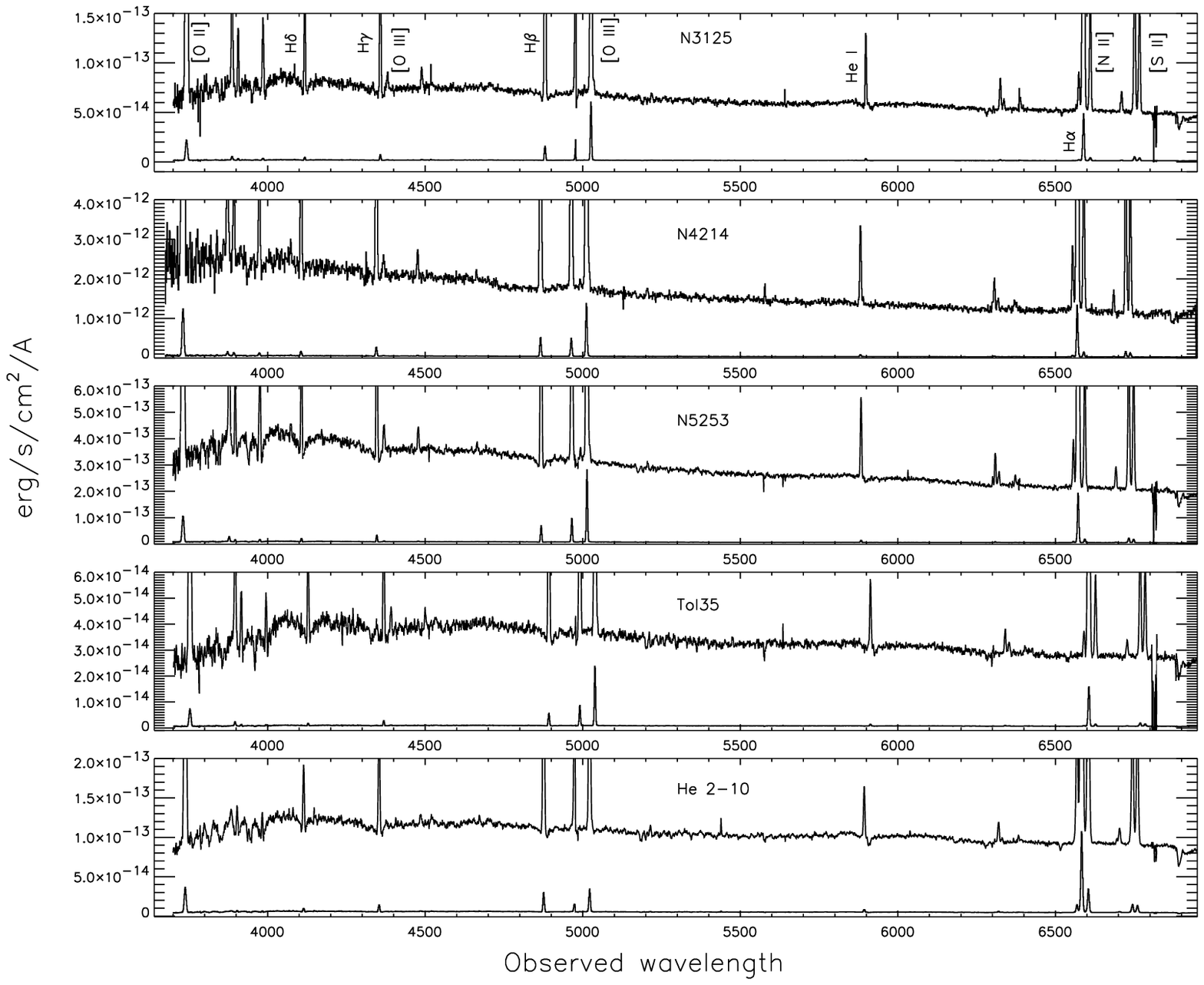,width=5.5in,angle=0}}
\figcaption[f1.ps] {Global spectra for each galaxy observed in this
work, not corrected for reddening or redshift.  Flux units are relative
units only, since the data were acquired using drift scan techniques.
Each spectrum is plotted a second time, scaled by a factor of 40 for
display purposes.  \label{spectra} }
\end{figure}

\begin{figure}
\centerline{\psfig{file=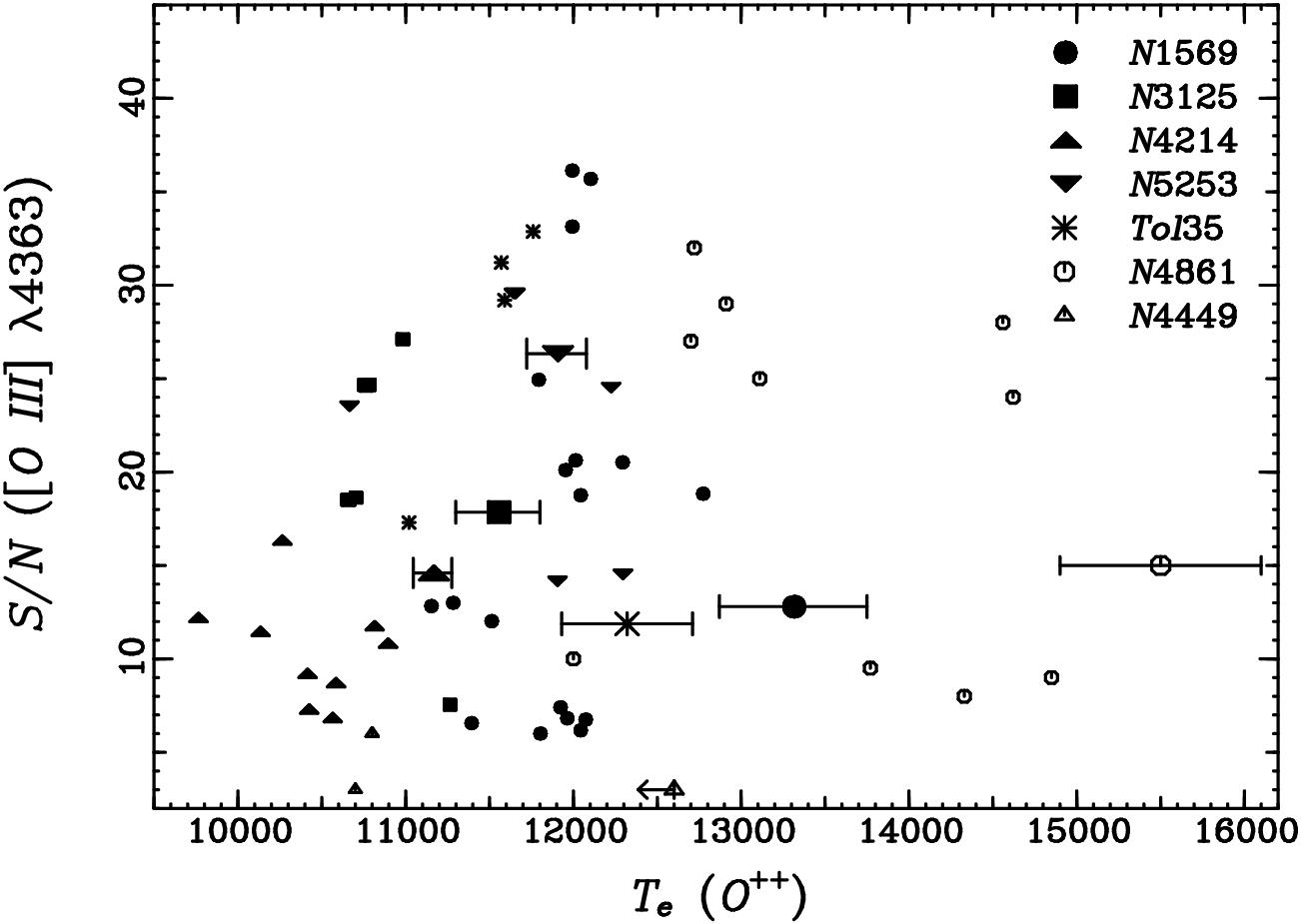,width=5.5in,angle=-90}}
\figcaption[f2.ps] {The electron temperature, $T_e$, derived from the
[O~III] $\lambda$4363 measurements versus the signal-to-noise ratio of
the [O~III] $\lambda$4363 line for multiple measurements in six nearby
irregular galaxies.  Small symbols represent measurements of individual
\HII\ regions or localized measurements, while large symbols with error
bars designate the results from global spectra.  For most objects,
[O~III] $\lambda$4363 in the global spectrum has a S/N ratio that lies
in the middle of the S/N distribution of small apertures.  The electron
temperature measured from the global spectra consistently fall near the
highest electron temperatures measured from localized spectra.  This is
consistent with global nebular galaxy spectra being biased toward
regions of highest temperature and surface brightness.  \label{Te_SN4363} }
\end{figure}

\begin{figure}
\centerline{\psfig{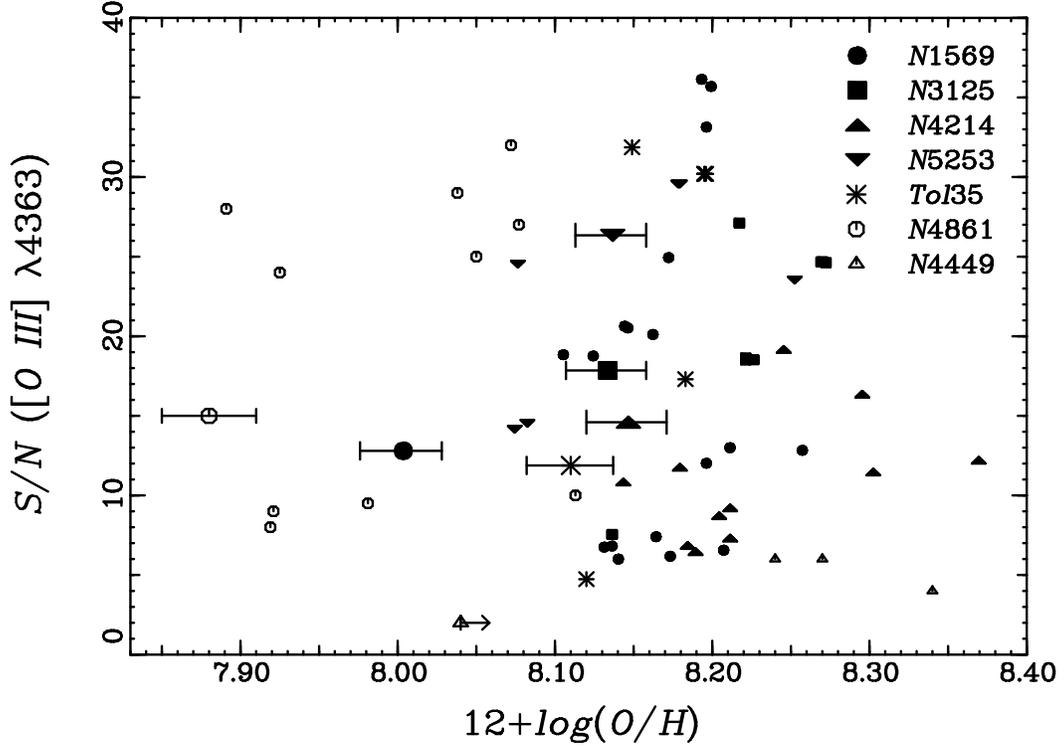}}
\figcaption[f3.ps] {The oxygen abundance, $12+{\log}(O/H)$, derived from
the [O~III] $\lambda$4363 measurements versus the signal-to-noise ratio
of the [O~III] $\lambda$4363 line for multiple measurements in six
nearby irregular galaxies.  Small symbols represent measurements of
individual \HII\ regions or localized measurements, while large symbols
with error bars designate the results from global spectra.  For most
objects, [O~III] $\lambda$4363 in the global spectrum has a S/N ratio
that lies in the middle of the S/N distribution of small apertures.
The but the oxygen abundances measured from the global spectra
consistently fall among the lowest values measured from localized
spectra.  This is consistent with global nebular galaxy spectra being
biased toward regions of highest temperature and surface brightness.
\label{OH_SN4363} }
\end{figure}

\begin{figure}
\centerline{\psfig{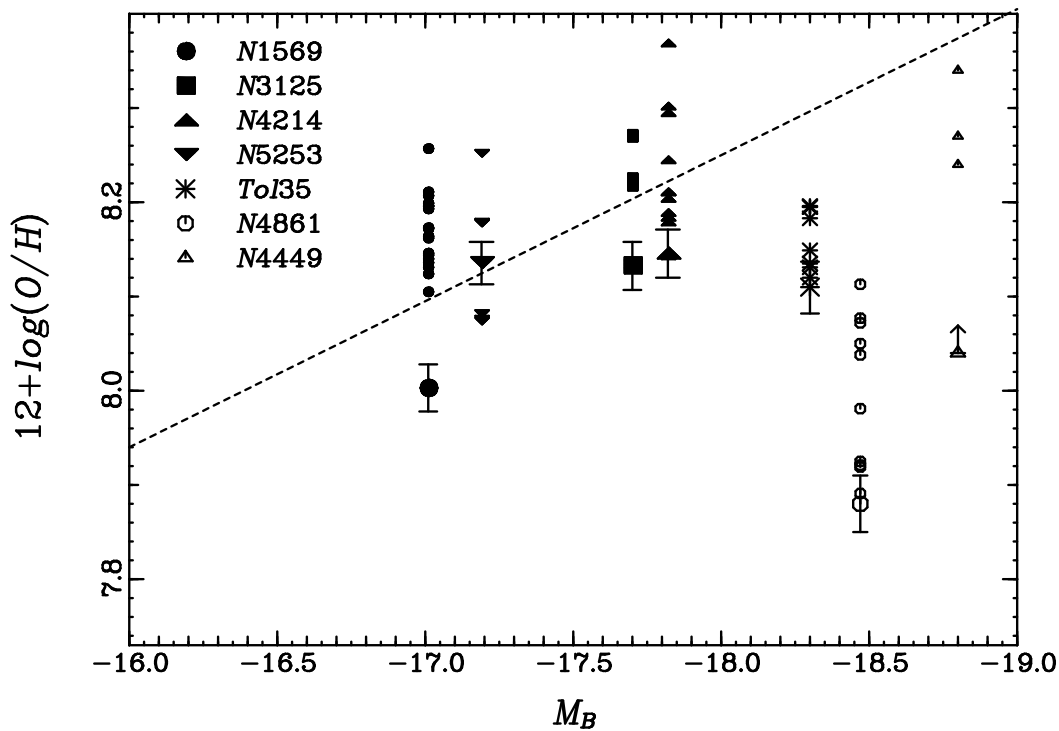}}
\figcaption[f4.ps] {Absolute blue magnitude versus oxygen abundance,
$12+{\log}(O/H)$, for six nearby irregular galaxies.  Small symbols
represent measurements of individual \HII\ regions or localized
measurements, while large symbols with error bars designate the results
from global spectra.  This figure demonstrates that the oxygen
abundances measured from the global spectra consistently fall among the
lowest values measured from localized spectra.  This is consistent with
global nebular galaxy spectra being biased toward regions of highest
temperature and surface brightness.    The data plotted here also show
the well-known correlation between luminosity and oxygen abundance for
irregular galaxies (cf. Skillman 1989)  \label{MB_OH} }
\end{figure}

\begin{figure}
\centerline{\psfig{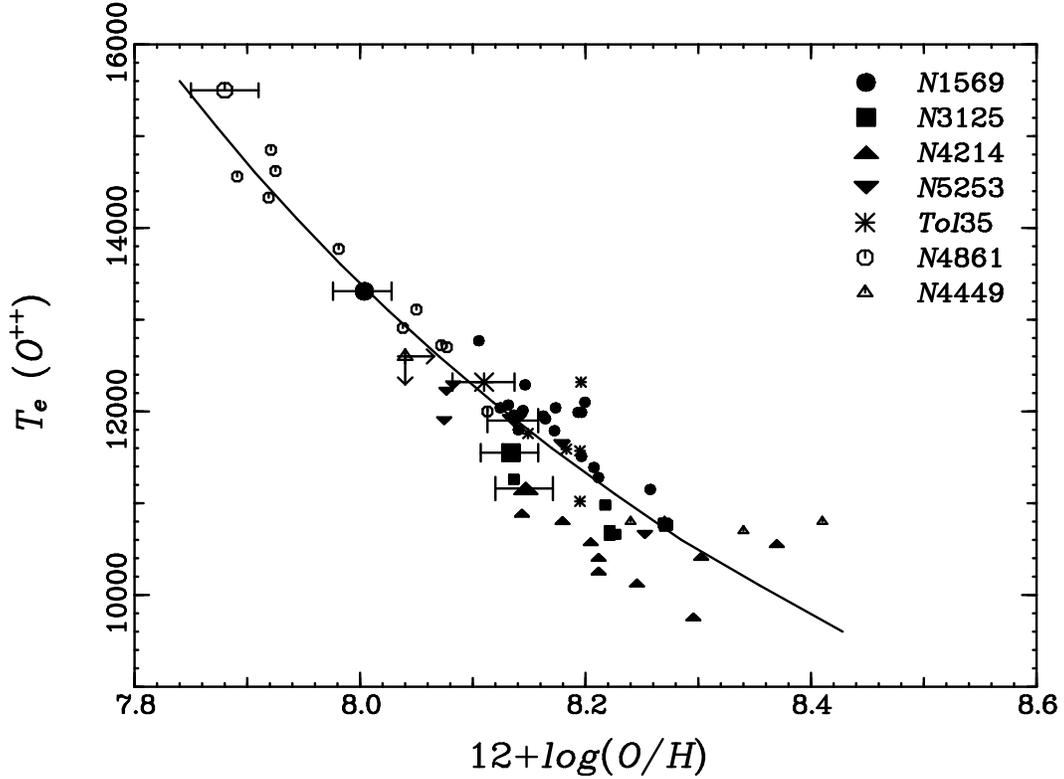}}
\figcaption[f5.ps]  {The [O~III] electron temperatures and the
resulting oxygen abundances for 69 positions in 6 metal-poor galaxies.
Small symbols represent measurements of individual \HII\ regions or
localized measurements, while large symbols denote global
measurements.  The solid line illustrates the direction along which O/H
measurements would scatter due to random uncertainties in the adopted
electron temperature.  Variations of 0.1--0.2 dex in O/H within
galaxies are consistent with variations in the adopted electron
temperature of 1000 K --- 2000 K.  The global measurement for each
galaxy consistently exhibits the largest $T_e$ and the smallest oxygen
abundance.  \label{OH_T} }
\end{figure}

\begin{figure}
\centerline{\psfig{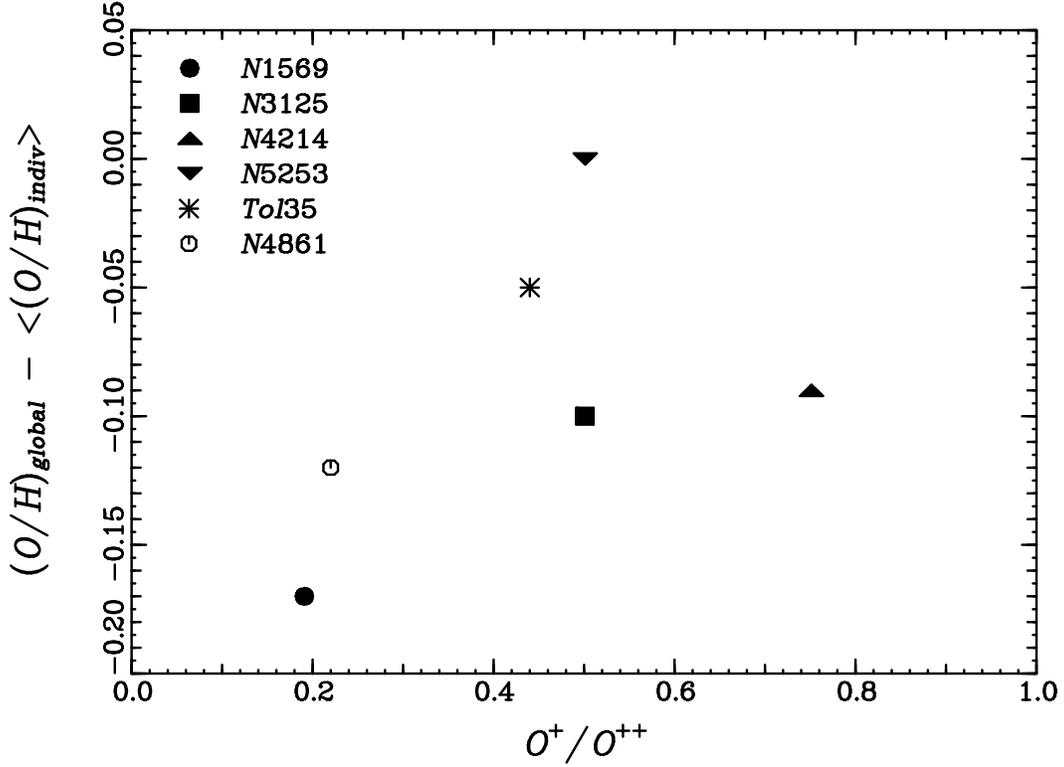}}
\figcaption[f6.ps]  {The ratio of singly to doubly ionized oxygen,
$O^+/O^{++}$, versus the difference between oxygen abundance derived
from global spectra and the mean of the spatially-resolved individual
measurements.  As seen in Figure~4, global spectra consistently yield
O/H measurements 0.05 --- 0.2 dex below the mean of the localized
measurements within the same galaxy.  There is a weak correlation,
consistent with the possibility that the global spectra in galaxies
dominated by doubly-ionized oxygen (i.e., those with $O^+/O^{++}<0.5$)
show a larger deviation from the mean measurement within that galaxy
compared to galaxies where $O^+$ and $O^{++}$ contribute more equally.
In none of the galaxies does $O^+$ dominate the nebular emission.  The
magnitude of the offset suggests a systematic error of $\sim0.1$ dex
when using global spectra to derive oxygen abundances from measured
electron temperatures and line strengths.  \label{O+O++_deltaOH} }
\end{figure}

\begin{figure}
\centerline{\psfig{file=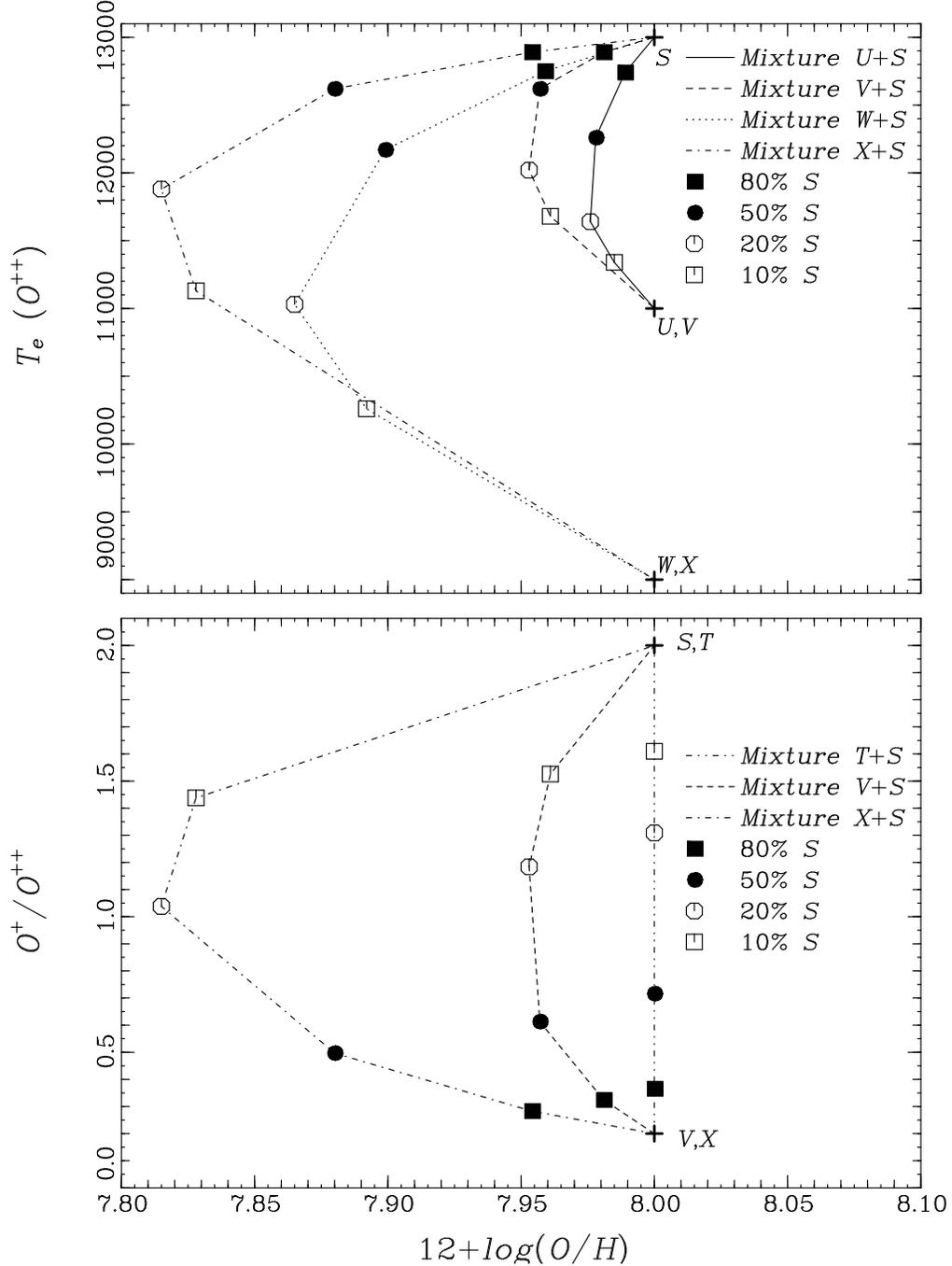,width=5.5in,angle=0}}
\figcaption[f7.ps] {Diagnostic diagrams showing how the measured oxygen
abundance, electron temperature, and ionization parameter (represented
by $O^+/O^{++}$) changes as nebulae with different $T_e$ and
$O^+/O^{++}$ are mixed in various proportions.  Crosses mark the
physical conditions of the basis spectra, S,T,U,V,W,X (see
Table~3).  Line styles distinguish gaseous mixtures of the standard
spectrum, $S$, ($T_e(O^{++})=13,000$ K, $O^+/O^{++}=0.2$), with each of
the other spectra.  Symbols trace the decreasing contribution of the
standard spectrum, $S$.  The top panel shows that even small amounts
of a high-temperature nebular component systematically bias derived
electron temperatures (oxygen abundances) to higher (lower) values.
For realistic temperature fluctuations and ionization variations, O/H
is underestimated by 0.05 to 0.2 dex.  \label{T-Udiag} }
\end{figure}

\begin{figure}
\centerline{\psfig{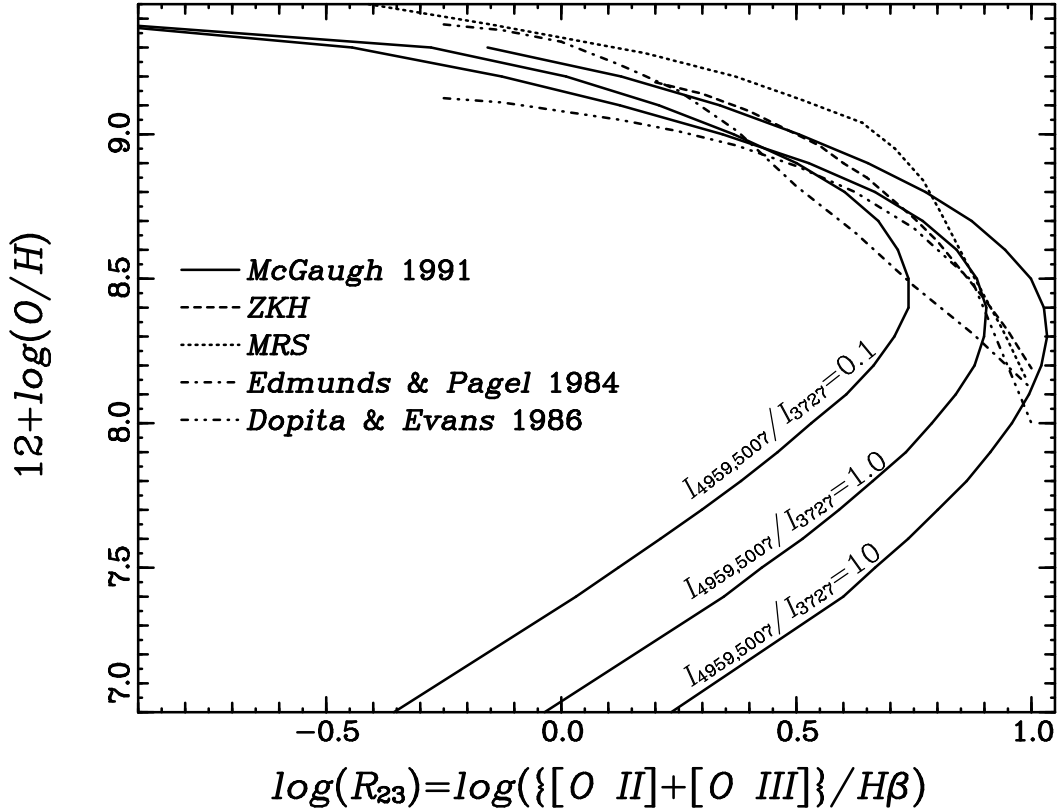}}
\figcaption[f8.ps] {The calibration of oxygen abundance, 12+log(O/H),
as a function of the strong line ratio ${\log}(R_{23})\equiv
\log([I_{3727}+I_{4959}+I_{5007}]/H\beta$) from several different
authors.  Strong line ratios can reliably indicate the oxygen abundance
to within the accuracy of the model calibrations, $\pm0.15$ dex.
However, the relationship is double-valued, requiring some $a~priori$
knowledge of a galaxy's metallicity in order to determine its correct
location on the upper or lower branch of the curve.    At low
metallicities, a given value of $R_{23}$ may yield different oxygen
abundances depending on the ionization parameter of the gas.  We show
the effect of varying the ionization parameter using the models of
McGaugh (1991) in terms of the observable line ratio,
$(I_{4959}+I_{5007})/I_{3727}$.    The uncertainties using this
empirical calibration are $\pm0.15$ dex, but larger, $\pm$0.25 dex n
the turn-around region near 12+log(O/H)$\sim$8.4.) \label{R23_OH.ps} }
\end{figure}

\begin{figure}
\centerline{\psfig{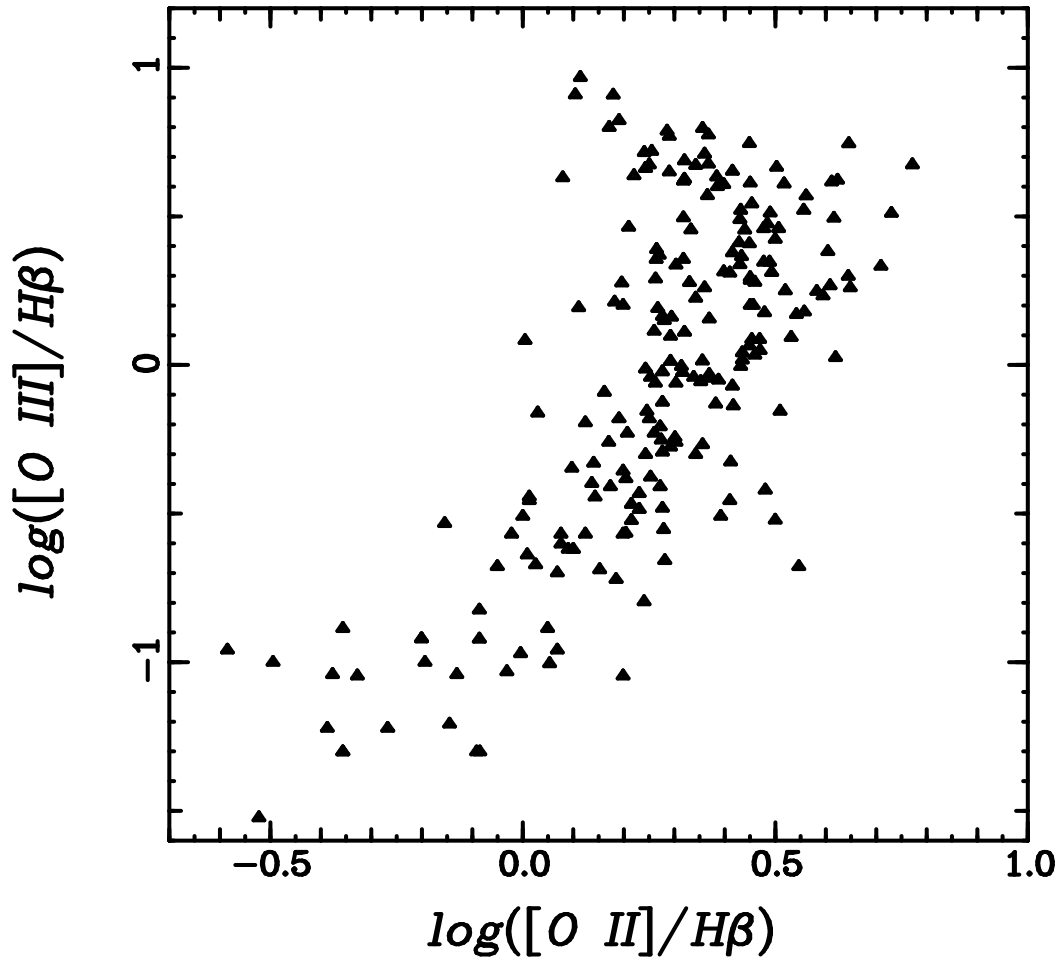}}
\figcaption[f9.ps] {log([O~III]/H$\beta$) versus log([O~II]/H$\beta$)
for \HII\ regions in 22 spiral galaxies from ZKH and
the literature.  Triangles
denote the global emission line spectrum for each galaxy constructed
from the sum of its constituent \HII\ region spectra (small symbols).
The global spectra follow the same excitation/ metallicity sequence as
individual \HII\ regions.  Labels at each end of the sequence denote
approximate metallicities of 2 $Z_\odot$ and 0.3 $Z_\odot$.
\label{R2_R3.ps}}
\end{figure}

\begin{figure}
\centerline{\psfig{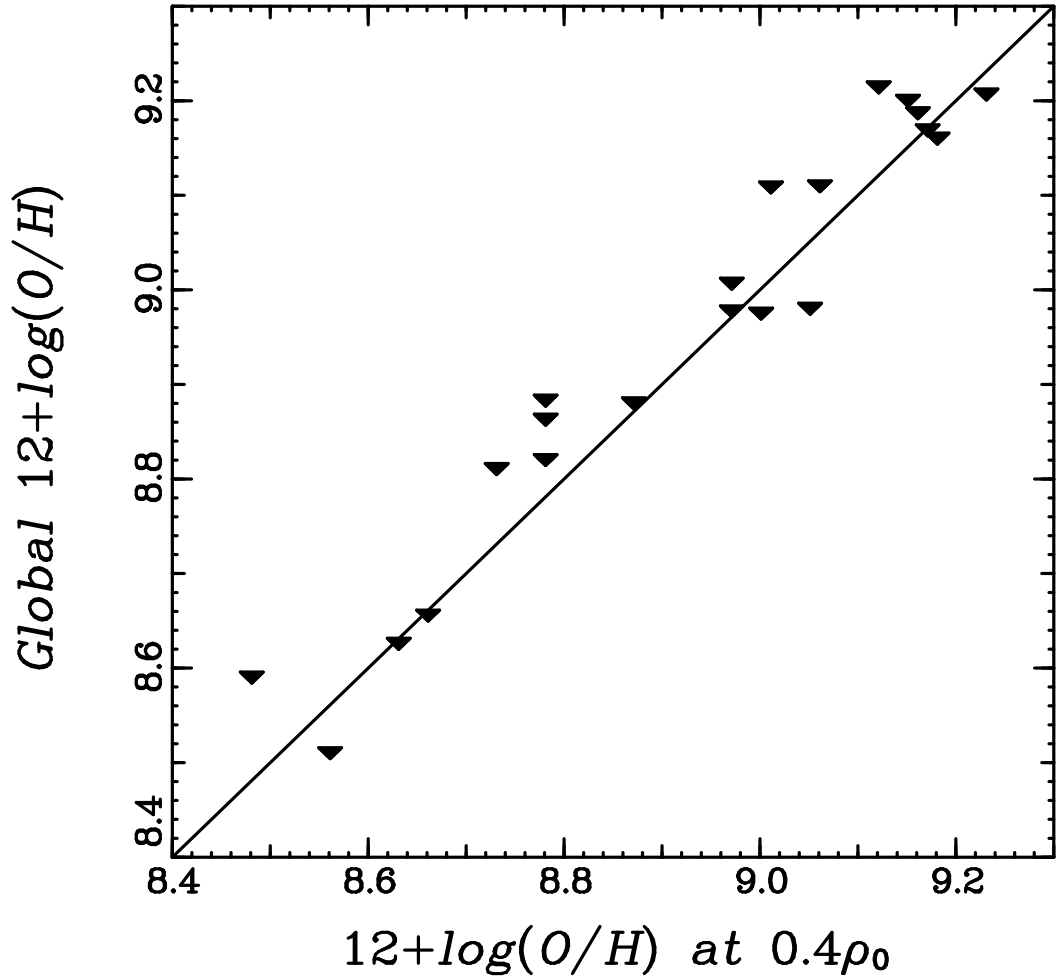}}
\figcaption[f10.ps] {The characteristic O/H ratio at a fiducial radius
of 0.4 isophotal radii as tabulated by ZKH versus the O/H ratio derived
from the global nebular spectrum for each galaxy. The global spectra
produce oxygen abundances which are in excellent agreement (0.1 dex)
with the value at 0.4 isophotal radii (25.0 mag arcsec$^{-2}$ in $B$).
\label{ZKH0.4_int.ps} }
\end{figure}


\begin{references}
\reference{} Aller, L.~H. 1942, ApJ, 95, 52
\reference{} Aller, L.~H. 1990, PASP, 102, 1097
\reference{} Brodie, J.~P., \& Huchra, J.~P. 1991, ApJ, 379, 157
\reference{} Dopita, M.~A., \& Evans, I.~N. 1986, ApJ, 307, 431
\reference{} Edmunds, M.~G. \& Pagel, B.~E.~J. 1984, MNRAS, 211, 507
\reference{} Faber, S.~M. 1973, ApJ, 179, 423
\reference{} Ferguson, A.~M., Wyse, R.~G., Gallagher, J.~S. III, \&
	Hunter, D.~A. 1996, AJ, 111, 2265
\reference{} French, H.~B., 1980, ApJ, 240, 41
\reference{} Garnett, D.~R. 1990, ApJ, 363, 142
\reference{} Gonzalez-Delgado, R.~M., P\'rez, Enrique, Tenorio-Tagle,
	G.,  \etal\ 1994, ApJ, 437, 239 
\reference{} Heckman, T.~M. 1980, A\&A, 87, 152
\reference{} Hummer, D.~G., \& Storey, P.~J. 1987, MNRAS, 224, 801
\reference{} Hunter, D.~A. 1994,  AJ, 107, 565
\reference{} Hunter, D.~A. \& Gallagher, J.~S. III 1990, ApJ, 362, 480
\reference{} Hunter, D.~A., \& Gallagher, J.~S. III. 1997, 475, 65
\reference{} Izotov, Y., Thuan, T.~T., \& Lipovetsky, V.~A. 1994, ApJ, 435, 647
\reference{} Kennicutt, R.~C. Jr. 1992, ApJS, 79, 255 (K92)
\reference{} Kennicutt, R.~C. Jr., \& Garnett, D.~R. 1996, ApJ, 456, 504
\reference{} Kingdon, J.~B., \& Ferland, G.~J. 1995, ApJ, 691
\reference{} Kobulnicky, H.~A. in Abundance Profiles: Diagnostic Tools for Galaxy History, ASP
   Conf. Ser. Vol. 147, eds. D. Friedli, M. Edmunds, C. Robert, \&
   L. Drissen (San Francisco: ASP)
\reference{} Kobulnicky, H.~A., \& Skillman, E.~D. 1996, ApJ, 471, 211 
\reference{} Kobulnicky, H.~A., \& Skillman, E.~D. 1997, ApJ, 489, 636
\reference{} Kobulnicky, H.~A. \& Zaritsky, D. 1998, ApJ, in press
\reference{}Lauroesch, J.~T., Truran, J.~W., Welty, D.~E., \& York, D.~G.
	1996, PASP, 108, 641
\reference{} Lequeux, J., Peimbert, M., Rayo, J.~F., Serrano, A., \& 
   Torres--Peimbert, S. 1979, A\&A, 80, 155
\reference{} Lilly, S.~J., Le F\'evre, O., Hammer, F., \& Crampton, D. 1996, 
	ApJ, 460, L1
\reference{} Liu, X., \& Danziger, J. 1995, MNRAS, 263, 256
\reference{} Madau, P. Ferguson, H.~C., Dickinson, M.~E., Giavalisco, M., Steidel, C.~C., \& Fruchter, A.
	1996, MNRAS, 283, 1388
\reference{} Martin, C.~L. 1996, ApJ, 465, 680
\reference{} Martin, C.~L. 1997, ApJ, 491, 561
\reference{} Martin, C.~L., \& Kennicutt, R.~C. 1998, ApJ, in press
\reference{} Martin, P., \& Roy, J.~R. 1994, 424, 599
\reference{} McCall, M.~L., Rybski, P.~M., \& Shields, G.~A. 1985,
        ApJS, 57, 1 (MRS)
\reference{} McGaugh, S. 1991, ApJ, 380, 140
\reference{} McGaugh, S. 1998, private communication
\reference{} Oke, J.~B. 1990, AJ, 99, 1621
\reference{} Olofsson, K. 1995, A\&AS, 111, 570
\reference{} Osterbrock, D.~E. 1989, { Astrophysics of Gaseous Nebulae
and Active Galactic Nuclei}, University Science Books:Mill Valley CA,
p. 134
\reference{} Pagel, B.~E.~J. 1986, PASP, 98, 1009
\reference{} Pagel, B.~E.~J., Edmunds, M.~G., Blackwell, D.~E., Chun, M.~S., \& Smith, G. 
	1979, MNRAS, 189, 95
\reference{} Pagel, B.~E.~J., Simonson, E.~A., Terlevich, R.~J., \& Edmunds, 
    M.~G. 1992, MNRAS, 255, 325
\reference{} Peimbert, M. 1967, ApJ, 150, 825 
\reference{} Peimbert, M. 1975, ARA\&A, 13, 113
\reference{} Peimbert, M. 1996,
	in The Analysis of Emission Lines, Poster Papers from the Space 
	Telescope Science Institute Symposium in Honor of the 70th 
	Birthdays of D. E. Osterbrock and M. J. Seaton, ed. R.~E. 
	Williams \&  M. Livio, (Baltimore: STScI), 165 
\reference{} P\'erez, E. 1997, MNRAS, 290, 465
\reference{} Richer, M.~G., \& McCall, M.~L. 1995, ApJ, 445, 642
\reference{} Searle, L. 1971, ApJ, 168, 327
\reference{} Shields, G.~A. 1990, ARA\&A, 28, 525
\reference{}  Skillman, E.~D., Bomans, D.~J., \& Kobulnicky, H.~A.
	1996, ApJ,  474, 205
\reference{} Skillman, E.~D., \& Kennicutt, R.~C. 1993, ApJ, 411, 655
\reference{}  Skillman, E.~D., Kennicutt, R.~C., \& Hodge, P. 1989,
	ApJ, 347, 875 (SKH)
\reference{} Stasi\'nska, G. 1990, A\&AS, 83, 501
\reference{} Steidel, C.~C., Giavalisco, M., Pettini, M., Dickinson, M.,
  Adelberger, K.~L. 1996, ApJ, 462, 17
\reference{} Steigman, G., Viegas, S.~M., \& Gruenwald, R. 1997, ApJ,
	490, 187.
\reference{} Talent, D.~L. 1980, Ph.D. Thesis, Rice University
\reference{} Veilleux, S., \& Osterbrock, D.~E. 1987, ApJS, 63, 295
\reference{} Villa-Costas, M.~B., \& Edmunds, M.~G. 1992, MNRAS, 259, 121
\reference{} Walsh, J.~R., \& Roy, J-R. 1989, MNRAS, 239, 297 
\reference{} Walsh, J.~R., \& Roy, J-R. 1993, MNRAS, 262, 27
\reference{} Walter, D.~K., Dufour, R.~J., \& Hester, J.~J. 1992, ApJ,
	397, 196
\reference{} Zaritsky, D., Kennicutt, R.~C., \& Huchra, J.~P. 1994, ApJ, 420,
	87 (ZKH)
\end{references}
\end{document}


\begin{planotable}{lrrrrrrrrr}
\tablecaption{Dereddened Global Line Strengths\tablenotemark{abc}   \label{Craw} }
\tablewidth{0pt}
\tablehead{
\colhead{Line} & 
\colhead{F($\lambda$)} & 
\colhead{NGC 1569}   &
\colhead{NGC 3125}   &
\colhead{NGC 4214}   & 
\colhead{NGC 4449}   & 
\colhead{NGC 4861}   & 
\colhead{NGC 5253}   & 
\colhead{Tololo 35}   &
\colhead{He 2-10}   \\ [.2ex]  
\colhead{}   & 
\colhead{}   & 
\colhead{${I_0}\over{I_{H\beta}}$ }   & 
\colhead{${I_0}\over{I_{H\beta}}$ }   & 
\colhead{${I_0}\over{I_{H\beta}}$ }   & 
\colhead{${I_0}\over{I_{H\beta}}$ }   & 
\colhead{${I_0}\over{I_{H\beta}}$ }   & 
\colhead{${I_0}\over{I_{H\beta}}$ }   & 
\colhead{${I_0}\over{I_{H\beta}}$ }   & 
\colhead{${I_0}\over{I_{H\beta}}$ }    }
\startdata
3727 [O \II]  & 0.25 & 0.988$\pm$0.066 & 2.440$\pm$0.078 & 3.038$\pm$0.089 & 3.891$\pm$0.257 & 1.248$\pm$0.092 & 2.597$\pm$0.101 & 2.410$\pm$0.079 & 2.289$\pm$0.073 \nl  
3868 [Ne \III]& 0.23 & 0.203$\pm$0.014 & 0.263$\pm$0.008 & 0.237$\pm$0.007 & 0.189$\pm$0.01  & 0.413$\pm$0.029 & 0.317$\pm$0.012 & 0.374$\pm$0.015 & $<$0.042 \nl  
4101 H$\delta$& 0.19 & 0.247$\pm$0.013 & 0.282$\pm$0.008 & 0.255$\pm$0.006 & 0.279$\pm$0.016 & 0.205$\pm$0.014 & 0.250$\pm$0.009 & 0.253$\pm$0.010 & 0.280$\pm$0.008 \nl  
4340 H$\gamma$& 0.12 & 0.425$\pm$0.015 & 0.459$\pm$0.011 & 0.490$\pm$0.010 & 0.492$\pm$0.017 & 0.487$\pm$0.027 & 0.446$\pm$0.015 & 0.453$\pm$0.012 & 0.452$\pm$0.011 \nl  
4363 [O \III] & 0.12 & 0.063$\pm$0.005 & 0.033$\pm$0.001 & 0.027$\pm$0.001 & $<$0.023        & 0.121$\pm$0.008 & 0.046$\pm$0.002 & 0.055$\pm$0.005 & $<$0.013 \nl  
4471 He \I    & 0.09 & 0.016$\pm$0.005 & 0.035$\pm$0.001 & 0.035$\pm$0.001 & $<$0.023        & 0.041$\pm$0.005 & 0.034$\pm$0.001 & 0.030$\pm$0.004 & 0.034$\pm$0.001 \nl  
4686 He \II   & 0.05 & 0.014$\pm$0.003\tablenotemark{d} & $<$0.007\tablenotemark{d} & 
	$<$0.006\tablenotemark{d} & $<$0.017\tablenotemark{d} & $<$0.004\tablenotemark{d} & 
	$<$0.007\tablenotemark{d} & $<$0.015\tablenotemark{d}  & $<$0.001  \nl  
4861 H$\beta$ &  0.00 & 1.000$\pm$0.003 & 1.000$\pm$0.022 & 1.000$\pm$0.018  & 1.000$\pm$0.006 & 1.000$\pm$0.004 & 1.000$\pm$0.032 & 1.000$\pm$0.023 & 1.000$\pm$0.022 \nl  
4969 [O \III] & -0.02 & 1.500$\pm$0.010 & ...\tablenotemark{e} & 1.004$\pm$0.018  & 0.689$\pm$0.007 & 1.954$\pm$0.093 & 1.421$\pm$0.045 & 1.539$\pm$0.035 & 0.361$\pm$0.008 \nl  
5007 [O \III] & -0.04 & 4.514$\pm$0.041 & 3.830$\pm$0.087 & 3.032$\pm$0.056  & 2.069$\pm$0.020 & 5.860$\pm$0.281 & 4.223$\pm$0.134 & 4.545$\pm$0.104 & 1.063$\pm$0.024 \nl  
5876 He \I    & -0.23 & 0.103$\pm$0.007 & 0.103$\pm$0.003 & 0.108$\pm$0.003  & 0.079$\pm$0.007 & 0.076$\pm$0.006 & 0.105$\pm$0.004 & 0.119$\pm$0.004 & 0.088$\pm$0.003 \nl  
6312 [S \III] & -0.30 & 0.016$\pm$0.004 & 0.014$\pm$0.001 & 0.014$\pm$0.001  & 0.019$\pm$0.005 & $<$0.003        & 0.017$\pm$0.001 & 0.024$\pm$0.002 & 0.006$\pm$0.000 \nl  
6548 [N \II]  & -0.34 & 0.042$\pm$0.004 & 0.056$\pm$0.002 & 0.082$\pm$0.003  & 0.115$\pm$0.011 & ...\tablenotemark{f} & 0.064$\pm$0.003 & 0.046$\pm$0.003 & 0.320$\pm$0.012 \nl  
6564 H$\alpha$& -0.34 & 2.840$\pm$0.252 & 2.841$\pm$0.110 & 2.855$\pm$0.103  & 2.866$\pm$0.254 & 2.792$\pm$0.256 & 2.828$\pm$0.126 & 2.849$\pm$0.111 & 3.002$\pm$0.116 \nl  
6584 [N \II] &  -0.34 & 0.137$\pm$0.012 & 0.183$\pm$0.007 & 0.241$\pm$0.009  & 0.338$\pm$0.031 & ...\tablenotemark{f} & 0.200$\pm$0.009 & 0.147$\pm$0.006 & 0.878$\pm$0.034  \nl  
6678 He \I   &  -0.35 & 0.027$\pm$0.003 & 0.032$\pm$0.001 & 0.028$\pm$0.001  & 0.028$\pm$0.006 & 0.050$\pm$0.006 & 0.032$\pm$0.002 & 0.031$\pm$0.002 & 0.028$\pm$0.001  \nl  
6717 [S \II] &  -0.36 & 0.205$\pm$0.019 & 0.246$\pm$0.010 & 0.267$\pm$0.010  & 0.476$\pm$0.045 & 0.174$\pm$0.017 & 0.270$\pm$0.012 & 0.243$\pm$0.010 & 0.301$\pm$0.012  \nl
6731 [S \II] &  -0.36 & 0.147$\pm$0.014 & 0.179$\pm$0.007 & 0.194$\pm$0.007  & 0.334$\pm$0.032 & 0.122$\pm$0.012 & 0.205$\pm$0.009 & 0.177$\pm$0.008 & 0.266$\pm$0.011  \nl  
\nl
Abs(H$\beta$)  (\AA)          &  & 6.0           & 3.0             & 1.0           & 1.5           & 2.0          & 2.0           & 1.5  & 2.0 \nl  
c(H$\beta$)                   &  & 0.75$\pm$0.05 & 0.29$\pm$0.04   & 0.19$\pm$0.04 & 0.16$\pm$0.03 & 0.18$\pm$0.10 & 0.20$\pm$0.04 & 0.31$\pm$0.04 & 0.54$\pm$0.04 \nl  
EW(H$\beta$)                   & & 62.3$\pm$0.5    & 49.5$\pm$1.1   & 58.4$\pm$1.1    & 11.6$\pm$  0.2  & 108.4$\pm$ 1.2    & 43.2$\pm$1.4  & 28.9$\pm$  0.5 & 28.2$\pm$  0.6   \nl  
n$_e$ (cm$^{-3}$)              & & $\leq$100.    & $\leq$100.     & $\leq$100     & $\leq$ 100.   & $\leq$100     &  $\leq$100    & $\leq$100 & 246$\pm$58 \nl  
\enddata
\tablenotetext{a}{Deredenned relative to H$\beta$ using a Galactic (Seaton 1986) redenning law.}
\tablenotetext{b}{All upper limits denote 3$\sigma$ upper limits.}
\tablenotetext{c}{Line strengths for NGC~1569, NGC~4449, and NGC~4861 are measured from
the global spectra presented in Kennicutt 1992.  NGC~3125, 5253, and Tololo 35 
line strengths are based on new spectra presented here.  A global spectrum for NGC~4214 is constructed from
the data reported in Kobulnicky \& Skillman (1996).}
\tablenotetext{d}{Includes only narrow, presumably nebular He~II.}
\tablenotetext{e}{The [O~III] $\lambda$4959 line measurement is unreliable due to defective columns in the CCD. }
\tablenotetext{f}{The spectra have insufficient resolution to measure [N~II] $\lambda\lambda$6548,6584 
	in such close proximity to H$\alpha$. }
\end{planotable}
\clearpage

\begin{planotable}{lrrrrrr}
\tablecaption{Derived Parameters and Abundances from Global Spectra   \label{phys} }
\tablewidth{0pt}
\tablehead{
\colhead{}   &
\colhead{$T(O^{++})$}   &
\colhead{$T(O^{+})$}   & 
\colhead{$O^+/H^+$ }   & 
\colhead{$O^{++}/H^+$ }   & 
\colhead{12+log(O/H) }   & 
\colhead{$log(N^+/O^+)$}  \\ [.2ex]  
\colhead{ Object }   & 
\colhead{ (K) }   & 
\colhead{ (K) }   & 
\colhead{($\times10^{-5}$)}   & 
\colhead{($\times10^{-5}$)}   & 
\colhead{}   & 
\colhead{}    }
\startdata
N1569      & 13010$\pm$440 & 12750$\pm$670 & 1.46$\pm$0.19 & 7.47$\pm$0.53 & 8.00$\pm$0.03  & -0.99$\pm$0.07   \nl  
N3125      & 11550$\pm$140 & 12010$\pm$520 & 4.50$\pm$0.57 & 9.00$\pm$0.31 & 8.13$\pm$0.02   & -1.30$\pm$0.06 \nl  
N4214      & 11160$\pm$150 & 11790$\pm$510 & 5.99$\pm$0.77 & 7.92$\pm$0.25 & 8.15$\pm$0.03   & -1.29$\pm$0.06 \nl  
N4449      & $<$12600      & $<$12300      & $>$6.60       & $>$4.26       & $>$8.05         & $<$-1.22       \nl  
N4861      & 15440$\pm$565 & 13820$\pm$750 & 1.41$\pm$0.18 & 6.17$\pm$0.50 & 7.88$\pm$0.03 & ...              \nl  
N5253      & 11900$\pm$200 & 12200$\pm$530 & 4.52$\pm$0.57 & 9.05$\pm$0.42 & 8.14$\pm$0.02 & -1.28$\pm$0.06   \nl  
Tololo 35  & 12320$\pm$390 & 12390$\pm$610 & 3.93$\pm$0.49 & 8.79$\pm$0.64 & 8.11$\pm$0.02 & -1.37$\pm$0.05  \nl  
Henize 2-10& $<$12100      & $<$12200      & $>$3.92       & $>$2.21       & $>$7.79 (8.93)\tablenotemark{a}  & $<-$0.57  \nl  
\enddata
\tablenotetext{a}{Oxygen abundance derived from strong line ratios
using the empirical relation of Zaritsky, Kennicutt, \& Huchra (1994) assuming
that Henize 2-10 lies on the upper branch of the R$_{23}$ calibration.}
\end{planotable}

\clearpage

\begin{planotable}{crrr}
\tablewidth{0pt}
\tablecaption{Parameters of Model Nebular Spectra \label{Obs} }
\tablehead{
\colhead{Mixture} & 
\colhead{12+log(O/H)\tablenotemark{a}}   & 
\colhead{$T_e (O~III)$\tablenotemark{b}}  & 
\colhead{$O^+/O^{++}$\tablenotemark{c} }    } 
\startdata
S  & 8.00 & 13000 & 0.20  \nl
T  & 8.00 & 13000 & 2.00  \nl
U  & 8.00 & 11000 & 0.20  \nl
V  & 8.00 & 11000 & 2.00  \nl
W  & 8.00 &  9000 & 0.20  \nl
X  & 8.00 &  9000 & 2.00  \nl
\enddata 
\tablenotetext{a}{Oxygen abundances assuming O/H = $O^+/H^+ + O^{++}/H^+$ }
\tablenotetext{b}{$O^{++}$ electron temperature as derived from
the [O~III] $\lambda$4363 and [O~III] $\lambda\lambda$ 4959,5007 line
ratios.}
\tablenotetext{c}{Ratio of singly to doubly ionized oxygen, by number,
as an indicator of the ionization parameter.}
\end{planotable}
\clearpage

\begin{planotable}{lccccccc}
\tablewidth{0pt}
\tablecaption{Dereddened Global Line Ratios and Oxygen Abundances\tablenotemark{a} \label{Obs} }
\tablehead{
\colhead{Galaxy} &
\colhead{log$([O~II]/H\beta)$} &
\colhead{log($[O~III]/H\beta)$} & 
\colhead{log$R_{23}$}   & 
\colhead{12+log(O/H)}  &
\colhead{} &
\colhead{} &
\colhead{}  \\ [.2ex]
\colhead{} &
\colhead{} &
\colhead{} &
\colhead{} &
\colhead{global} &
\colhead{at 0.4$\rho_0$\tablenotemark{b} } &
\colhead{at 3 kpc\tablenotemark{b}} &
\colhead{at 0.8$\rho_s$\tablenotemark{b} } }
\startdata
NGC~598  & 0.32 & 0.36 & 0.64 & 8.86 & 8.78 & 8.70 & 8.91 \nl
NGC~925  & 0.49 & 0.51 & 0.80 & 8.63 & 8.63 & 8.68 & 8.68 \nl
NGC~2403 & 0.45 & 0.29 & 0.68 & 8.81 & 8.73 & 8.80 & 8.82 \nl
NGC~2541 & 0.29 & 0.65 & 0.81 & 8.61 & 8.48 & 8.45 & ...  \nl
NGC~2903 & 0.01 & $-$0.46 & 0.14 & 9.21 & 9.12 & 9.22 & 9.31 \nl
NGC~3031 & 0.27 & 0.19 & 0.53 & 8.97 & 9.00 & 8.89 & 9.10 \nl
NGC~3184\tablenotemark{c} & 0.14 & $-$0.33 & 0.27 & 9.16 & 9.18 & 9.26 & ... \nl
NGC~3198 & 0.41 & 0.31 & 0.67 & 8.82 & 8.78 & 8.94 & 8.98 \nl
NGC~3319 & 0.36 & 0.71 & 0.87 & 8.49 & 8.56 & 8.63 & 8.65 \nl
NGC~3344\tablenotemark{c} & 0.27 & 0.37 & 0.62 & 8.88 & 8.87 & 8.75 & 9.17 \nl
NGC~3351 & $-$0.09 & $-$0.82 & -0.01 & 9.27 & 9.31 & 9.33 & ... \nl
NGC~3521 & 0.35 & $-$0.06 & 0.50 & 9.00 & 8.97 & 9.08 & 9.40 \nl
NGC~4254 & 0.14 & $-$0.44 & 0.24 & 9.17 & 9.17 & 9.26 & 9.27 \nl
NGC~4258 & 0.32 & 0.11 & 0.53 & 8.98 & 8.97 & 9.07 & 9.06 \nl
NGC~4303 & 0.16 & $-$0.09 & 0.35 & 9.11 & 9.01 & 9.20 & 9.25 \nl
NGC~4321 & 0.07 & $-$0.60 & 0.16 & 9.21 & 9.23 & 9.32 & 9.28 \nl
NGC~4559 & 0.51 & 0.46 & 0.78 & 8.66 & 8.66 & 8.79 & ...  \nl
NGC~4725\tablenotemark{c} & 0.10 & $-$0.62 & 0.17 & 9.20 & 9.15 & 9.26 & ... \nl
NGC~5033 & 0.28 & 0.15 & 0.52 & 8.98 & 8.84 & 9.09 & 8.96 \nl
NGC~5236 & 0.12 & $-$57 & 0.20 & 9.18 & 9.16 & 9.17 & 9.19 \nl
NGC~5457 & 0.30 & 0.33 & 0.62 & 8.88 & 8.52 & 8.76 & 8.78 \nl
NGC~6946 & 0.24 & $-$0.30 & 0.35 & 9.12 & 9.06 & 9.13 & 9.10 \nl
\enddata 
\tablenotetext{a}{The global emission line characteristics
derived from individual \HII\ regions spectra
which have been weighted according
to the radial emission line luminosity as described in the text.}
\tablenotetext{b}{The mean azimuthally averaged oxygen abundance
at 0.4 of the isophotal radius (25.0 mag arcsec$^{-2}$ in $B$), 0.8 of the
disk scale length, and 3 kpc radius, taken
from Zaritsky, Kennicutt, \& Huchra (1994).}
\tablenotetext{c}{Global $R_{23}$ values and global O/H are
computed from the direct sum of \HII\ region spectra
from ZKH, rather than from a radial weighted
emission-line intensity since emission line images
were not available.  It is remarkable that even the 
unweighted global spectra closely reproduce the
mean abundance at 0.4$\rho_0$, to within 0.05 dex.}
\end{planotable}
\clearpage